\documentclass[dvipsnames, twocolumn, twocolappendix]{aastex63}
\usepackage[version=4]{mhchem}
\usepackage{amsmath}

\received{XX}
\revised{YY}
\accepted{ZZ}
\submitjournal{ApJ}

\shorttitle{MAPS XV: Tracing structure within 20 au}
\shortauthors{Bosman \& the MAPS collaboration}

\begin{document}
\title{Molecules with ALMA at Planet-forming Scales (MAPS). XV.

Tracing protoplanetary disk structure within 20 au}
\date{today}

\correspondingauthor{Arthur Bosman}
\email{arbos@umich.edu}

\author[0000-0003-4001-3589]{Arthur D. Bosman}
\affiliation{Department of Astronomy, University of Michigan, 323 West Hall, 1085 S. University Avenue, Ann Arbor, MI 48109, USA}

\author[0000-0003-4179-6394]{Edwin A. Bergin}
\affiliation{Department of Astronomy, University of Michigan,
323 West Hall, 1085 S. University Avenue,
Ann Arbor, MI 48109, USA}

\author[0000-0002-8932-1219]{Ryan A. Loomis}
\affiliation{National Radio Astronomy Observatory, Charlottesville, VA 22903, USA}

\author[0000-0003-2253-2270]{Sean M. Andrews}
\affiliation{Center for Astrophysics \textbar\ Harvard \& Smithsonian, 60 Garden St., Cambridge, MA 02138, USA}

\author[0000-0002-2555-9869]{Merel L.R. van \'{}t Hoff}
\affiliation{Department of Astronomy, University of Michigan, 323 West Hall, 1085 S. University Avenue, Ann Arbor, MI 48109, USA}

\author[0000-0003-1534-5186]{Richard Teague}
\affiliation{Center for Astrophysics \textbar\ Harvard \& Smithsonian, 60 Garden St., Cambridge, MA 02138, USA}

\author[0000-0001-8798-1347]{Karin I. \"Oberg}
\affiliation{Center for Astrophysics \textbar\ Harvard \& Smithsonian, 60 Garden St., Cambridge, MA 02138, USA}

\author[0000-0003-4784-3040]{Viviana V. Guzm\'an}
\affil{Instituto de Astrof\'isica, Pontificia Universidad Cat\'olica de Chile, Av. Vicu\~na Mackenna 4860, 7820436 Macul, Santiago, Chile}

\author[0000-0001-6078-786X]{Catherine Walsh}
\affiliation{School of Physics and Astronomy, University of Leeds, Leeds LS2 9JT, UK}

\author[0000-0003-3283-6884]{Yuri Aikawa}
\affiliation{Department of Astronomy, Graduate School of Science, The University of Tokyo, Tokyo 113-0033, Japan}

\author[0000-0002-2692-7862]{Felipe Alarc\'on}
\affiliation{Department of Astronomy, University of Michigan,
323 West Hall, 1085 S. University Avenue,
Ann Arbor, MI 48109, USA}

\author[0000-0001-7258-770X]{Jaehan Bae}
\altaffiliation{NASA Hubble Fellowship Program Sagan Fellow}
\affil{Earth and Planets Laboratory, Carnegie Institution for Science, 5241 Broad Branch Road NW, Washington, DC 20015, USA}

\author[0000-0002-8716-0482]{Jennifer B. Bergner}
\altaffiliation{NASA Hubble Fellowship Program Sagan Fellow}
\affiliation{University of Chicago Department of the Geophysical Sciences, Chicago, IL 60637, USA}

\author[0000-0003-2014-2121]{Alice S. Booth}
\affiliation{Leiden Observatory, Leiden University, 2300 RA Leiden, the Netherlands}
\affiliation{School of Physics and Astronomy, University of Leeds, Leeds, LS2 9JT UK}

\author[0000-0002-2700-9676]{Gianni Cataldi}
\affil{National Astronomical Observatory of Japan, Osawa 2-21-1, Mitaka, Tokyo 181-8588, Japan}
\affil{Department of Astronomy, Graduate School of Science, The University of Tokyo, Tokyo 113-0033, Japan}

\author[0000-0003-2076-8001]{L. Ilsedore Cleeves}
\affiliation{University of Virginia, 530 McCormick Rd, Charlottesville, VA 22904, USA}

\author[0000-0002-1483-8811]{Ian Czekala}
\altaffiliation{NASA Hubble Fellowship Program Sagan Fellow}
\affiliation{Department of Astronomy and Astrophysics, 525 Davey Laboratory, The Pennsylvania State University, University Park, PA 16802, USA}
\affiliation{Center for Exoplanets and Habitable Worlds, 525 Davey Laboratory, The Pennsylvania State University, University Park, PA 16802, USA}
\affiliation{Center for Astrostatistics, 525 Davey Laboratory, The Pennsylvania State University, University Park, PA 16802, USA}
\affiliation{Institute for Computational \& Data Sciences, The Pennsylvania State University, University Park, PA 16802, USA}
\affiliation{Department of Astronomy, 501 Campbell Hall, University of California, Berkeley, CA 94720-3411, USA}

\author[0000-0001-6947-6072]{Jane Huang}
\altaffiliation{NASA Hubble Fellowship Program Sagan Fellow}
\affiliation{Department of Astronomy, University of Michigan, 323 West Hall, 1085 S. University Avenue, Ann Arbor, MI 48109, USA}
\affiliation{Center for Astrophysics \textbar\ Harvard \& Smithsonian, 60 Garden St., Cambridge, MA 02138, USA}

\author[0000-0003-1008-1142]{John D. Ilee}
\affiliation{School of Physics and Astronomy, University of Leeds, Leeds LS2 9JT, UK}

\author[0000-0003-1413-1776]{Charles J. Law}
\affiliation{Center for Astrophysics \textbar\ Harvard \& Smithsonian, 60 Garden St., Cambridge, MA 02138, USA}

\author[0000-0003-1837-3772]{Romane Le Gal}
\affiliation{Center for Astrophysics \textbar\ Harvard \& Smithsonian, 60 Garden St., Cambridge, MA 02138, USA}
\affiliation{IRAP, Universit\'{e} de Toulouse, CNRS, CNES, UT3, 31400 Toulouse, France}
\affiliation{Univ. Grenoble Alpes, CNRS, IPAG, F-38000 Grenoble, France}
\affiliation{IRAM, 300 rue de la piscine, F-38406 Saint-Martin d'H\`{e}res, France}

\author[0000-0002-7616-666X]{Yao Liu}
\affiliation{Purple Mountain Observatory \& Key Laboratory for Radio Astronomy, Chinese Academy of Sciences, Nanjing 210023, China}

\author[0000-0002-7607-719X]{Feng Long}
\affiliation{Center for Astrophysics \textbar\ Harvard \& Smithsonian, 60 Garden St., Cambridge, MA 02138, USA}

\author[0000-0002-1637-7393]{Fran\c cois M\'enard}
\affiliation{Univ. Grenoble Alpes, CNRS, IPAG, F-38000 Grenoble, France}

\author[0000-0002-7058-7682]{Hideko Nomura}
\affiliation{National Astronomical Observatory of Japan, Osawa 2-21-1, Mitaka, Tokyo 181-8588, Japan}

\author[0000-0002-1199-9564]{Laura M. P\'erez}
\affiliation{Departamento de Astronomía, Universidad de Chile, Camino El Observatorio 1515, Las Condes, Santiago, Chile}

\author[0000-0001-8642-1786]{Chunhua Qi}
\affiliation{Center for Astrophysics \textbar\ Harvard \& Smithsonian, 60 Garden St., Cambridge, MA 02138, USA}

\author[0000-0002-6429-9457]{Kamber R. Schwarz}
\altaffiliation{NASA Hubble Fellowship Program Sagan Fellow}
\affiliation{Lunar and Planetary Laboratory, University of Arizona, 1629 East University Boulevard, Tucson, AZ 85721, USA}

\author[0000-0002-5991-8073]{Anibal Sierra}
\affiliation{Departamento de Astronomía, Universidad de Chile, Camino El Observatorio 1515, Las Condes, Santiago, Chile}

\author[0000-0002-6034-2892]{Takashi Tsukagoshi}
\affiliation{National Astronomical Observatory of Japan, Osawa 2-21-1, Mitaka, Tokyo 181-8588, Japan}

\author[0000-0003-4099-6941]{Yoshihide Yamato}
\affiliation{Department of Astronomy, Graduate School of Science, The University of Tokyo, Tokyo 113-0033, Japan}

\author[0000-0003-1526-7587]{David J. Wilner}
\affiliation{Center for Astrophysics \textbar\ Harvard \& Smithsonian, 60 Garden St., Cambridge, MA 02138, USA}

\author[0000-0002-0661-7517]{Ke Zhang}
\altaffiliation{NASA Hubble Fellow}
\affiliation{Department of Astronomy, University of Wisconsin-Madison,
475 N Charter St, Madison, WI 53706}
\affiliation{Department of Astronomy, University of Michigan,
323 West Hall, 1085 S. University Avenue,
Ann Arbor, MI 48109, USA}





\begin{abstract}
Constraining the distribution of gas and dust in the inner 20 au of protoplanetary disks is difficult. At the same time, this region is thought to be responsible for most planet formation, especially around the water ice line at 3-10 au. Under the assumption that the gas is in a Keplerian disk, we use the exquisite sensitivity of the Molecules with ALMA at Planet-forming Scales (MAPS) ALMA large program to construct radial surface brightness profiles with a $\sim$3 au effective resolution for the CO isotopologue $J=$2--1 lines using the line velocity profile.
IM Lup reveals a central depression in \ce{^{13}CO} and \ce{C^{18}O} that is ascribed to a pileup of $\sim$500 $M_\oplus$ of dust in the inner 20 au, leading to a gas-to-dust ratio of around $<$10. This pileup is consistent with efficient drift of grains ($\gtrsim$ 100 $M_\oplus$~Myr$^{-1}$) and a local gas-to-dust ratio that suggests that the streaming instability could be active. The CO isotopologue emission in the GM Aur disk is consistent with a small ($\sim$ 15 au), strongly depleted gas cavity within the $\sim$40 au dust cavity. The radial surface brightness profiles for both the AS 209 and HD 163296 disks show a local minimum and maximum in the \ce{C^{18}O} emission at the location of a known dust ring ($\sim$14 au) and gap ($\sim$10 au), respectively. This indicates that the dust ring has a low gas-to-dust ratio ($>$ 10) and that the dust gap is gas-rich enough to have optically thick \ce{C^{18}O}. This paper is part of the MAPS special issue of the Astrophysical Journal Supplement.
\end{abstract}


\section{Introduction}

Exoplanet statistical studies imply that the majority of solar-type stars have a planet within 1 au of the star, with observed close-in planets spanning a wide mass range, from sub-earth to multi-Jupiter-mass planets \citep[e.g.][]{Johnson2010, Mulders2018}. It is hypothesized that a large portion of these planets form relatively close to the star, with the water ice line posited as a  favored location to facilitate planet formation \citep[e.g.][]{Ciesla2006, Lyra2010,  Cridland2019, Fernandes2019}. This primary zone of planet formation is difficult to probe, as the physical scales are small, $<$10 au in radius, corresponding to angular sizes of $<0\farcs1$ in the closest star-forming regions (140--200 parsec).

The inner 20~au is also critical in terms of disk physics. It is the region where the magnetorotational instability is thought to be strongly suppressed and accretion toward the star is assumed to be powered by magnetohydrodynamic winds \citep[][]{Armitage2011}. Internal photoevaporative winds are also thought to be launched from the inner 20 au as well \citep[for a review, see][]{Ercolano2017}. Changes in transport speed of the gas, as well as the launching of a wind are thought to have profound effects on the structure of the gas disk, which then impacts the dust disk as well. Observing structure in the inner 20~au could thus reveal information on a host of processes.

Studies of this primary planet-forming zone have focused on either the composition of the molecular gas within the water ice line or the structure of the gas and dust disk near the dust sublimation radius. Compositional studies of the disk inside the water ice line, are driven by near- and mid-infrared observations, with, for example, VLT-CRIRES, Keck-NIRSPEC, VLT-VISIR and \textit{Spitzer}-IRS. The observations generally lack the resolution to spatially resolve the inner disk, but it is only the inner disk that has the physical conditions necessary to produce emission of the 2--35 $\mu$m rotational and ro-vibrational lines that these instruments target \citep[e.g.][]{Carr2008,Pontoppidan2010,Salyk2011}. These observations have taught us that the surface layers of the inner disk are hot (500 -- 1000~K) \citep[e.g.,][]{Salyk2011}, dust poor \citep[e.g.,][]{Meijerink2009}, and strongly UV irradiated \citep{Pontoppidan2014, Bosman2018}.

Directly imaging structure in the inner 10s of au is difficult as high resolution optical and infrared imagers either obscure this region behind a coronagraph or have the stellar PSF overwhelm the disk emission \citep[e.g.][]{Avenhaus2017}. Submillimeter interferometry can now reach 30 mas ($\sim$5 au) resolution in the dust, barely resolving this region \citep[e.g.][]{ALMA2015, Andrews2018}. However, gas emission line studies in protoplanetary disks are generally limited to 100 mas in light of sensitivity and integration time considerations \citep[e.g.][]{oberg20}. Infrared-interferometry can reach a resolution down to 1 mas in both the gas and dust, which would easily resolve the planet formation regions, however, these instruments generally are not able to probe scales larger than $\sim$5 au and are most sensitive to emission on scales smaller than 1 au \citep[][]{Dullemond2010, Menu2015, Lazareff2017, Gravity2017}. There is thus a gap in our knowledge of gas structure at few au scales in the inner $\sim$20 au from imaging studies.

Exploiting the spatial information in high resolving power spectra ($\frac{\lambda}{\Delta \lambda} = R>25000$) it is possible to close this gap in our knowledge for gas emission lines. This has mostly been applied to the strong infrared CO ro-vibrational lines around 4.7 $\mu$m \citep[e.g.,][]{Pontoppidan2008spectro, vanderPlas2015, Banzatti2015, Bosman2019}. In particular these observations have been used to map the CO column density profile in the dust cavity of transition disk HD 139614 \citep[][]{Carmona2017}.

\begin{table*}[!t]
{\small
    \caption{Source properties}
    \begin{tabular}{l c c c c r}
    
    \hline
    \hline
    Source & M$_\star$  & Incl& $v_\mathrm{sys}$& $v_\mathrm{range}$& Reference\\
    & ($M_\odot$) &  (deg) & (km s$^{-1}$)&  (km s$^{-1}$)& \\
    \hline
    IM Lup & 1.1 & 47.5 & 4.5 & $-$19.5, 28.5 &\citep{Pinte2018, Huang2018, czekala20}\\
    GM Aur & 1.1 & 53.2 & 5.6 & $-$22.4,33.6  &\citep{Macias2018, Huang2020}\\
    AS 209 & 1.2 & 35.0 & 4.6 & $-$29.4, 38.6 & \citep{Huang2017, Huang2018, czekala20}\\
    HD 163296 & 2.0 & 46.7 & 5.8& $-$33.0, 44.8 &\citep{Andrews2018, Huang2018, Teague2019}\\
    MWC 480 & 2.1 & 37.0 & 5.1 & $-$26.9, 37.1$^{a}$&\citep{Pietu2007, Simon2019, Liu2019}\\
    \hline
    \end{tabular}
    \tablecomments{$^{a}$ Available velocity range for \ce{^{13}CO} and \ce{C^{18}O} is only -6.9, 17.1 km s$^{-1}$}
    }
    \label{tab:mass_incl}
\end{table*}

The high resolving power of submillimeter interferometers ($R > 10^6$) also allows the use of kinematic information to extract spatial information. Notable results include the inference of significant molecular gas within the millimeter dust hole in TW Hya at radii $<4$ au \citep{Rosenfeld2012}, a gap in the molecular gas in GM Aur \citep{Dutrey2008} that has just recently been resolved \citep{Huang2020, law20_rad}. It has further been used to constrain the gas distribution in debris disks \citep{Hales2019} and the CO gas mass within an unresolved CO snowline in a handful of Class I and II sources \citep{Zhang2020a, Zhang2020b}.  

In this paper, we use the high sensitivity and spectral resolving power of the Molecules with ALMA at Planet-forming Scales (MAPS) data \citep{oberg20, czekala20, law20_rad} to zoom in on the CO emission in the inner few au of the targeted disks (AS 209, IM Lup, GM Aur, HD163296 and MWC 480). \citet{law20_rad} presents central flux depressions in the CO line emission in four out of five (all except MWC 480) of the MAPS sources in some or all of the isotopologue lines. The goals of this paper are to look for and, where possible, characterize and explain unresolved structure in the CO emission. We are thus tracing gas structure down to the primary planet-forming zone.

\section{Observations}

This study uses CO line data taken as part of the MAPS ALMA Large Program (2018.1.01055.L), specifically the \ce{^{12}CO}, \ce{^{13}CO} and, \ce{C^{18}O} $J$=2--1 and the \ce{^{13}CO} and \ce{C^{18}O} $J$=1--0 isotopologue lines.

The reduction and imaging procedure of these data is outlined in \citet{oberg20} and \citet{czekala20}\footnote{\url{http://www.alma-maps.info}}. From the standard data products we used the circularized, 0\farcs3 beam images for all isotopologue lines.
For the $J$=2--1 lines, we also use images with minor differences in the imaging procedure, namely velocity range imaged for these lines is 4x as large as detailed in \citet{czekala20} to get a proper baseline of line free channels (See Table~\ref{tab:mass_incl} for the velocity ranges). With this velocity range, image cubes with a 0\farcs15 and 0\farcs3 circularized beam are created for all three isotopologues. Furthermore, we use line+continuum CLEAN mask, which combine the Keplerian mask, with an elliptical mask that encompasses the millimeter disk, even though we are imaging the continuum-subtracted visibilities. The line+continuum CLEAN masks make sure that all on source flux in the high velocity channels is included. Unless otherwise noted, we use these wider CO images for our analysis.

For these wider velocity range images, as for the fiducial CO images (\ce{^{13}CO} and \ce{C^{18}O}, $J$=1--0; \ce{^{12}CO}, \ce{^{13}CO} and \ce{C^{18}O}, $J$=2--1) spectra are extracted from the image cubes by summing the pixels in either a circular aperture or in the CLEAN mask.

\section{Methods}
\label{sec:methods}
\begin{figure*}
    \centering
    \includegraphics[width=\hsize]{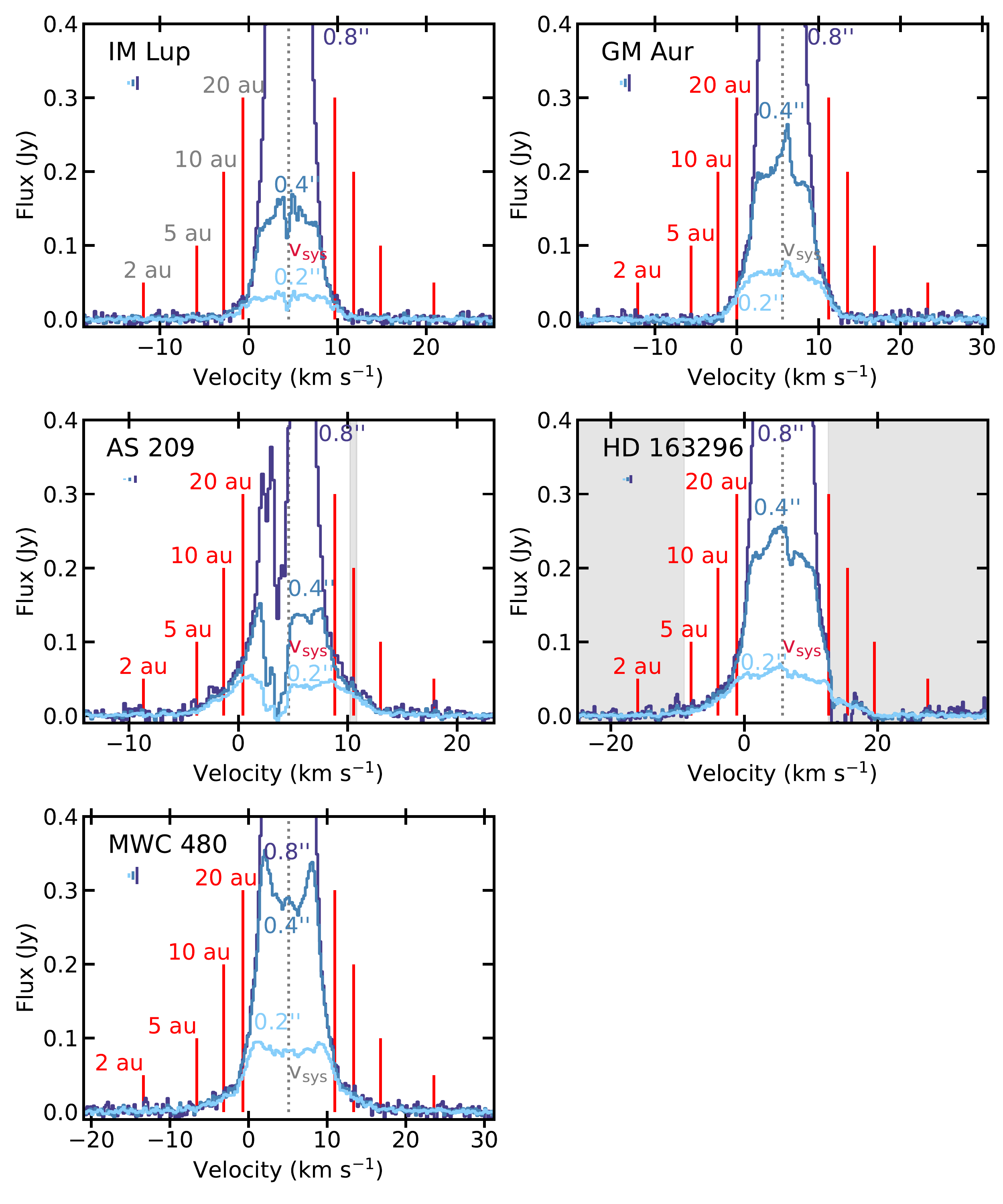}
    \caption{\ce{CO} $J$=2--1 spectrum of all our sources as extracted with different circular apertures. The apertures have a radius between 0\farcs2 and 0\farcs8. The vertical bars under the disk names show the error bars for the extracted spectra. The grey dotted vertical line shows the systemic velocity, red vertical lines show the velocity that corresponds to the maximal projected velocity for 2, 5, 10 and 20 au and the grey shaded areas show the regions that have been masked in the radial profile determination, these regions are discussed in Appendix~\ref{ssc:linesym} }
    \label{fig:All_apts}
\end{figure*}

Figure~\ref{fig:All_apts} shows the spectral line profiles extracted for the \ce{^{12}CO} $J$=2--1 line towards all five of the MAPS sources using a series of circular apertures with a radius, ranging between 0\farcs2 and 0\farcs8. The smallest aperture (0\farcs2) has the advantage that it includes fewest pixels, and that at large velocities (and thus small emitting radii) it is the most precise measurement. Comparison to larger apertures shows that around the velocities corresponding to 10-20 au (around 0\farcs15) the spectrum from the smallest aperture starts to deviate as it no longer encompasses all the flux in the image.

The symmetry of the line profiles has been studied in Appendix~\ref{ssc:linesym}, where we only find nonsymmetric emission in AS 209 and HD 163296, which will be masked in the rest of the analysis (see Fig.~\ref{fig:All_apts}). A comparison of the $J$=2--1 and $J$=1--0 line profiles is presented in Appendix~\ref{ssc:band_rat}. No clear evidence of the lower frequency line probing deeper into the disk, or significantly less impact of the continuum subtraction is seen in the comparison of the line profiles.

The spectrum used for the radial profile fitting is a combination of the spectra extracted with the different apertures. The spectra are extracted from the wide velocity range $J$=2--1 isotopologue images. For channels that have emission that should only originate within 0.05\arcsec, a 0\farcs4 aperture is used, and for emission originating between 0\farcs05 and 0\farcs25 a 0.6\arcsec{} aperture is used . For these images a baseline 0\farcs15 circularized beam is used. For emission originating between 0\farcs25 and 0\farcs5, the spectrum is extracted with a 0.8\arcsec{} aperture from the images that have a 0.3\arcsec{} circularized beam. As 0.5\arcsec{} translates to 50-80 au, this spectrum contains all the flux necessary for our purposes. This has been checked against the flux extracted from within the CLEAN mask. For any of our further analysis, we will not consider emission from radii $>40$ au. The velocity range in the spectra that corresponds to these radii is never taken into account in the fitting.

Flux errors are estimated by taking the RMS of the complete ALMA data cube outside the CLEAN mask and multiplying that by the square root of the number of beams that fit within the spectral extraction aperture.

The reconstruction of the radial intensity profiles ($I(R)$) is based on the assumption that all the disk emission is coming from gas in Keplerian rotation. For gas in Keplerian rotation, the maximal projected velocity that is achieved at a given radius is given by,
\begin{equation}
\label{eq:v_r_kepl}
v_\mathrm{max}(R) = \sqrt{\frac{G M_\star}{R}} \sin(i),
\end{equation}
where $G$ is Newton's gravitational constant, $M_\star$ is the stellar mass, $R$ the radius and $i$ the inclination of the disk. Stellar mass and inclinations used in this paper are listed in Table.~\ref{tab:mass_incl}. This implies that emission at velocity offsets larger than a given value, must be generated within the radius as given by Eq.~\ref{eq:v_r_kepl}. For each of the disks these relations are given in Fig.~\ref{fig:r_vs_v}.  This implies a relation between $\frac{\mathrm{d}F(v)}{\mathrm{d}v}$ and $I(R)$. While it is in principle possible to use this relation directly to derive $I(R)$ from the extracted spectra, we do not do this, but will instead use a very simple forward model to fit the spectra ($F(v)$) with a radial intensity profile. Fitting a forward model makes it easier to directly account for the finite velocity resolution and estimate the effect of noise in the spectra on the inferred brightness profiles. Full details of the fitting procedure are given in Appendix~\ref{app:fitting}.

\section{Radial intensity profiles}
\begin{figure}
    \centering
    \includegraphics[width=0.95\hsize]{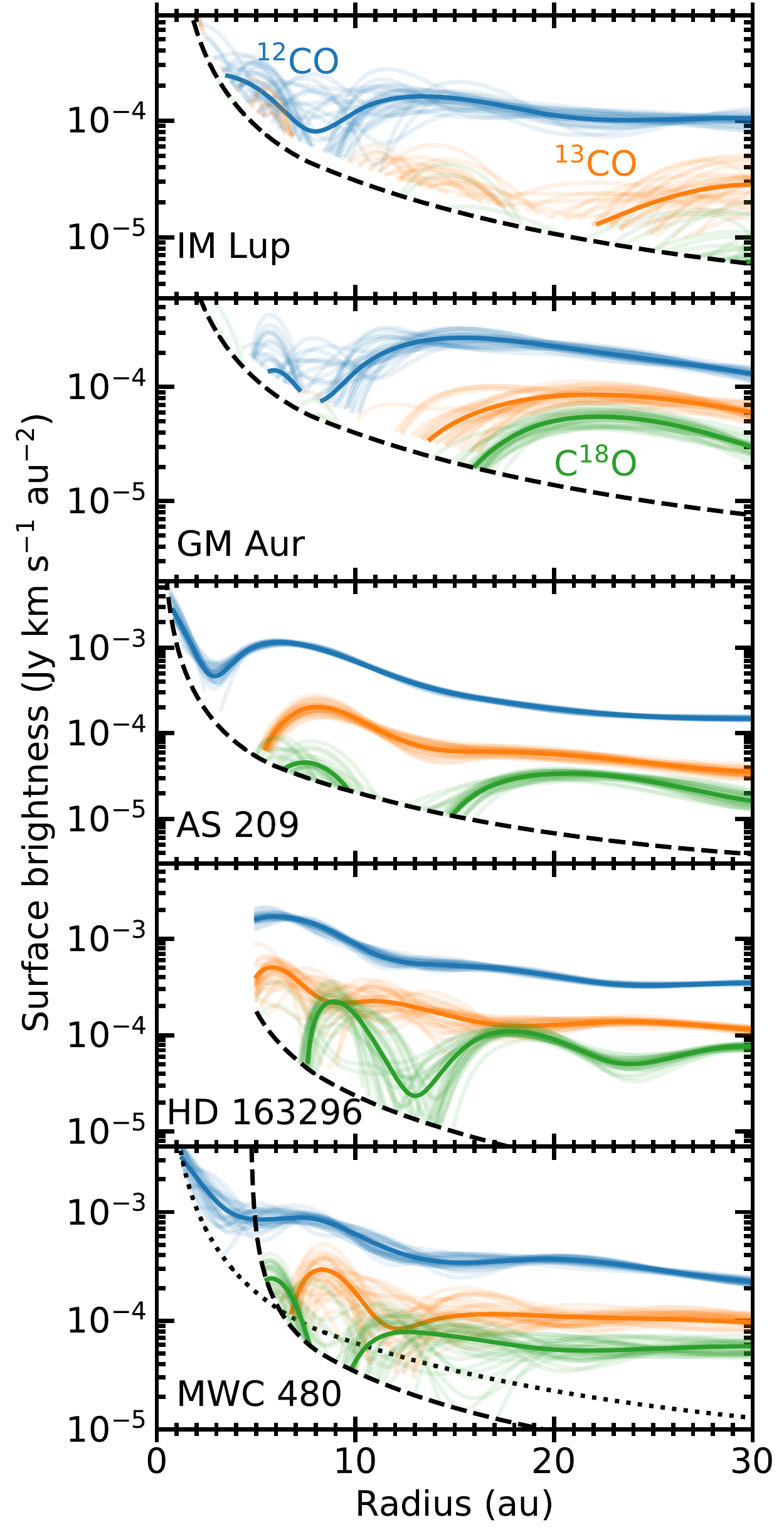}
    \caption{Radial profiles of the five MAPS sources as constrained from the \ce{^{12}CO} (blue), \ce{^{13}CO} (orange) and \ce{C^{18}O} (green) $J$=2--1 spectra. The fit to the data is shown in a thick line. Thin lines show 30 fits of the data after flux offsets had been applied to each velocity bin according to the observational uncertainties. These lines give an estimate on the uncertainty in the strength, depth and location of the features. The black dashed line shows the lower limit to the flux that can be measured at each radii for the \ce{C^{18}O} spectra. When a radial surface brightness profile drops below its detection limit, it is not plotted. Due to slightly different RMS values for the \ce{^{12}CO}, \ce{^{13}CO} and \ce{C^{18}O} data cubes this does not happen exactly at the dashed line for the \ce{^{12}CO} and \ce{^{13}CO} radial profiles. The inner radius for HD 163296 is $\sim$ 5 au as the highest velocities had to be masked out. For MWC 480 a wider velocity range was available for the \ce{^{12}CO} than the \ce{^{13}CO} and \ce{C^{18}O}, in this case the \ce{^{12}CO} sensitivity is given in a black dotted line.}
    \label{fig:Radial_profiles}
\end{figure}

Figure~\ref{fig:Radial_profiles} shows the radial surface brightness profiles extracted from the CO isotopologue line profiles for the five MAPS sources. These surface brightness profiles show features on scales that are not distinguishable in the CLEANED images with 0\farcs15 resolution.

The surface brightness of the \ce{^{13}CO} and \ce{C^{18}O} isotopologues are low in the inner regions of the IM Lup disk. The \ce{C^{18}O} surface brightness is below our detection threshold within 30 au and the \ce{^{13}CO} surface brightness drops below the detection threshold around 20 au. The \ce{^{12}CO} surface brightness profiles show some structure between 5 and 15 au, with a local minimum a that is a factor two lower than the surrounding surface brightness at $\sim$8 au. Given the large errors at these radii, it is assumed that these are driven by the noise in the data.

The radial profiles of the GM Aur disk all show a strong drop in emission in the inner region. \ce{^{12}CO} drops inside 15 au, while \ce{^{13}CO} and \ce{C^{18}O} drop inside of 20 au. This leads to very similar \ce{C^{18}O} and \ce{^{13}CO} profiles, with the \ce{^{13}CO} profile being slightly brighter.
Comparing to the radial profiles derived from the CLEANED images by \citet{law20_rad} (see~Appendix~\ref{app:image_comp}), the surface brightness derived from the line profiles shows a steeper drop in all isotopologues.

The radial surface brightness profiles of the AS 209 disk are rich in substructures. The \ce{^{12}CO} profile appears centrally peaked and has a local minimum at $\sim$3 au followed by a local maximum at $\sim$8 au with a monotonically decreasing flux towards larger radii. The \ce{^{13}CO} seems to follow the \ce{^{12}CO} radial profile relatively well showing the same peak at 8 au and dropping inward from this radius, after which the emission becomes indistinguishable from the noise. The \ce{C^{18}O} shows a strong decrease in surface brightness inward of 20 au that is not seen in the \ce{^{12}CO} and \ce{^{13}CO} lines. While at low significance, the \ce{C^{18}O} does also show a maximum at 8 au, together with the other isotopologues.

The surface brightness profiles of the HD 163296 disk only probe down to 5 au, as the velocity channels corresponding to smaller radii are contaminated on both the blue- and red-shifted side. The \ce{^{12}CO} and \ce{^{13}CO} surface brightness profiles show low amplitude (less than a factor 2) variations over the entire 5-30 au range. These amplitude variations do not seem to be consistent between the lines, however. The \ce{C^{18}O} shows strong (more than a factor 2) oscillations between 7 and 15 au. At around 9 au there is a peak in the surface brightness distribution, leading to comparable \ce{C^{18}O} and \ce{^{13}CO} fluxes. Within 9 au there is a sharp drop in surface brightness to below our detection threshold. Outside of 9 au the flux drops to a local minimum at 12 au.

The radial surface brightness profiles of \ce{^{12}CO} and \ce{^{13}CO} around MWC 480 show similar low amplitude deviations as HD 163296, with \ce{^{13}CO} additionally showing a strong peak around 10 au. The \ce{C^{18}O} surface brightness profile shows a strong dip at the same location. The \ce{^{12}CO} peaks strongly towards the inner disk.

\section{The effect of dust on line emission}

The inner regions of these five protoplanetary disks show a large variety of structures that is not well traced by imaging with the high, 0.15\arcsec, resolution of the MAPS data. Four of the disks, AS 209, IM Lup, HD 163296, and MWC 480, have strong dust emission in the inner 30 au. AS 209 and HD 163296 also have previously resolved structure in the dust disk in this region \citep{Andrews2018, Huang2018}. Before interpreting these structures, it is necessary to discuss the effects of dust on the line emission.

\begin{figure}
    \centering
    \includegraphics[width = \hsize]{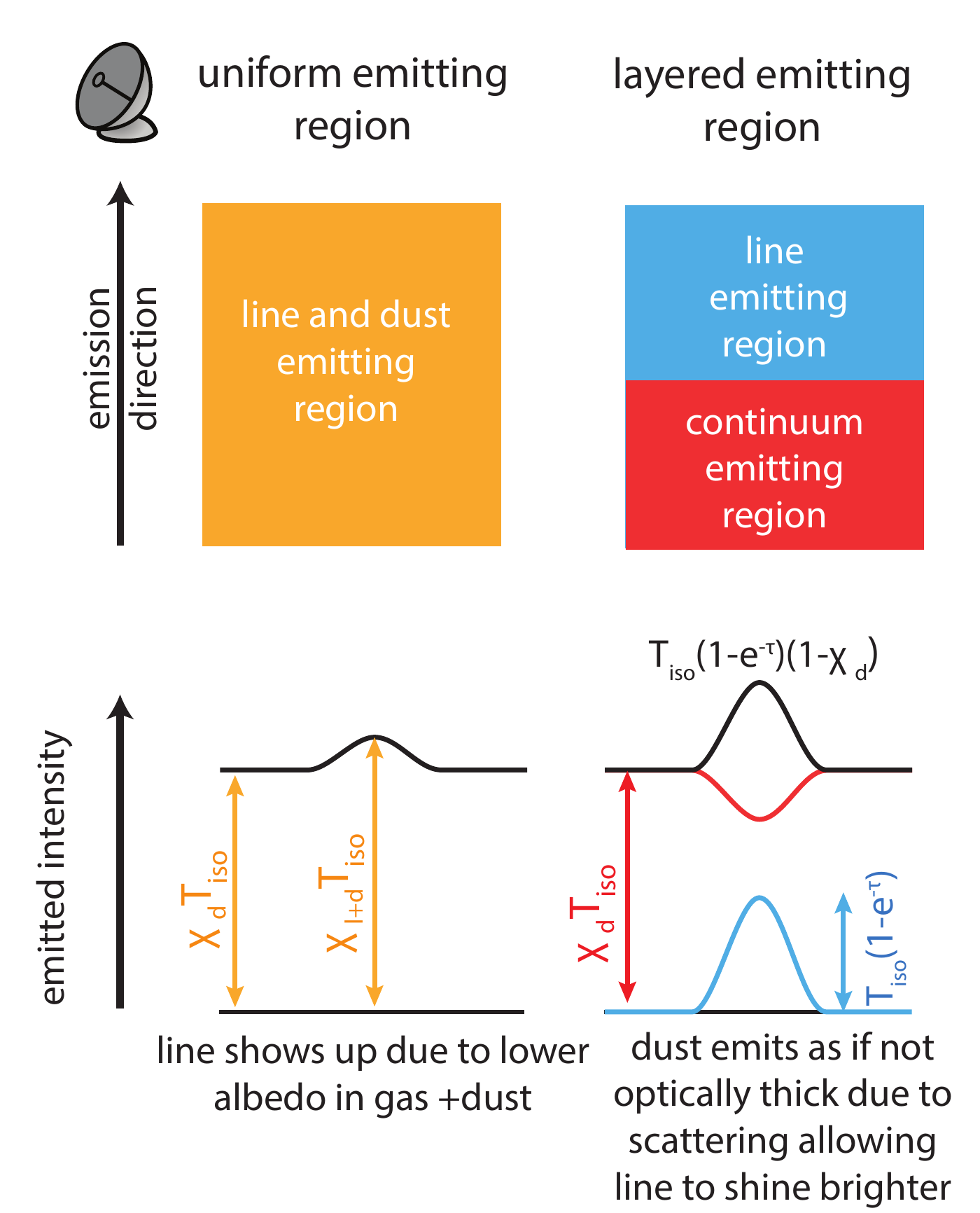}
    \caption{Schematic representation of the expected line strengths from an isothermal layer that has a optically thick dust layer and an optically thin gas contribution. The scenario in orange (left, Eq.~\ref{eq:I_scatsame}) shows an emission layer where gas and dust are well mixed. The scenario on the right (Eq.~\ref{eq:I_scatlayer}) depicts a molecular line emitting region (blue) on top of a dust (continuum) emitting region (red). In all cases the observer is assumed to look upon the layers from above as noted by the telescope dish.
    In the bottom panels the expected emission of this layer around a gas spectral line is shown, observed spectrum is shown in black in the right panel the continuum and gas contribution to the spectrum is shown in red and blue respectively.  $\chi_{\mathrm{d}}$ refers to the dust intensity reduction coefficient \citep[Eq.~\ref{eq:chi}, ][]{Zhu2019} and $\chi_{\mathrm{l+d}}$ the line and dust intensity reduction coefficient (Eq.~\ref{eq:child})}
    \label{fig:dust_line_iso}
\end{figure}

Given the limited velocity coverage of the CO spectral windows (100--200 km s$^{-1}$) which is on par with the infrared, rovibrational CO line widths \citep[e.g.][]{Brown2013, Banzatti2017}, a single spectral window cannot be used to conclusively distinguish high velocity (small radii) CO emission from continuum emission. All the spectra used here are thus continuum subtracted, using the dust emission information in all spectral windows combined for an accurate continuum determination.

The role of the dust optical depth on the CO emission profiles is not lessened in this way and standard checks, such as comparing with continuum or line+continuum images, are not possible. It is thus worth discussing the effects of (unresolved) continuum emission on the line radial profiles. As the dust emission of all the MAPS disks has been modeled \citep[][]{zhang20, sierra20}, there is a good estimate of the average millimeter dust optical depth in the inner few tens of au. The millimeter emission of GM Aur drops within 40 au, with multi-band observations implying that this emission is optically thin. AS 209 and HD 163296 appear to be optically thin around 220 GHz in dust emission down to 20 au assuming a non-scattering dust model \citep{zhang20,sierra20}, however the dust model including scattering, and the evidence of dust substructure within 20 au imply that dust is optically thick in some regions within 40 au \citep[][]{Andrews2018, Huang2018, Guzman2018, sierra20}. IM Lup and MWC 480 are consistent with fully optically thick emission within $\sim$ 40 au even without assuming strong scattering at millimeter wavelengths \citep[][]{Liu2019, zhang20, sierra20}.

Optically thick dust has several effects on the observed line emission. Aside from obscuring gas under the millimeter dust photo-sphere, it also produces a background flux for the line emission originating above the dust emission surface. When gas and dust temperatures differ, there will still be a line visible, either in emission or absorption and gas properties can still be extracted \citep[see, e.g.][]{Weaver2018}. There is a relevant edge case, however, when gas (or excitation temperature) and dust temperature are the same over the line of sight. This can happen in an vertically isothermal layer in the inner disk.

The surface brightness profiles of the IM Lup, AS 209, HD 163296 and MWC 480 disks showing drops of factors of 3 or more.  It is necessary to consider these surface brightness decline as the result of optically thick dust in an isothermal layer and investigate the exact conditions necessary to cause these strong surface brightness drops.

\subsection{Line emission with dust scattering}

To investigate this, we consider two very simplified physical configurations as shown in Fig.~\ref{fig:dust_line_iso}. We consider an isothermal layer of temperature $T_\mathrm{iso} = T_\mathrm{ex} = T_\mathrm{dust}$ that either has the line and dust emitting region completely overlapping, or vertically completely separated. For each of these configurations we derive an expression for the continuum subtracted line flux, and discuss the conditions under which the continuum-subtracted line emission is most efficiently diminished.

When looking at millimeter wavelengths, scattering by dust is not negligible. The scattering opacity at a wavelength of 1 mm from grains with a size distribution $n(a) \propto a^{p}$, with $p = -3.5$, between 0.005 $\mu$m and 1 mm is a factor 10 more than the absorption opacity \citep[][]{Birnstiel2018, zhang20}. This reduces the continuum flux up to a factor of 4 \citep[see][for a full discussion]{Zhu2019}. Following \citet{Zhu2019}, we write the optically thick dust intensity as $I_\mathrm{dust} = \chi B_{\nu}\left(T_\mathrm{iso}\right)$; $\chi$ is the intensity reduction coefficient, and is approximately given as \citep[see][Eq. 13]{Zhu2019}
\begin{equation}
\label{eq:chi}
   \chi \approx \sqrt{1-\omega}
   = \sqrt{1 - \kappa_\mathrm{scat}/\left(\kappa_\mathrm{abs} + \kappa_\mathrm{scat}\right)},
\end{equation}
where $\kappa_\mathrm{scat}$ and $\kappa_\mathrm{abs}$ are the scattering and absorption opacity and $\omega$ is the dust albedo. If line and dust are emitting from the same medium the line + continuum intensity reduction coefficient has to be used. This is given by:
\begin{equation}
\begin{aligned}
\label{eq:child}
   \chi_\mathrm{line+dust} \approx   \sqrt{1 - \frac{\kappa_\mathrm{scat, dust}}{\left(\kappa_\mathrm{abs, dust} + \kappa_\mathrm{abs, line} + \kappa_\mathrm{scat, dust}\right)}}.
\end{aligned}
\end{equation}
Note that the absorption of photons by molecules does not have a scattering component associated with it, as such $\chi_\mathrm{line+dust} > \chi_\mathrm{dust}$ always holds. Assuming that the dust optical depth $\tau_\mathrm{dust} \gg 1$, the continuum subtracted line intensity $I_\mathrm{L-C}$ is given by:
\begin{equation}
\begin{aligned}
\label{eq:I_scatsame}
       I_\mathrm{L-C} = &\chi_\mathrm{line+dust}B_\nu\left(T_\mathrm{iso}\right) -
        \chi_\mathrm{dust}B_\nu\left(T_\mathrm{dust}\right)\\
       = &B_\nu\left(T_\mathrm{iso}\right) \left(\chi_\mathrm{line+dust} - \chi_\mathrm{dust} \right),
\end{aligned}
\end{equation}
with $B_\nu(T)$ is the black body radiation. In this case (Fig.~\ref{fig:dust_line_iso}, left), the continuum subtracted line emission only disappears if the opacity of the line is negligible compared to the dust opacity, as that would result in $\chi_\mathrm{line+dust} \approx \chi_\mathrm{dust}$ (see Sec.~\ref{sss:linesuppression}, for the necessary conditions for CO).

In a protoplanetary disk, the line emission is generated over a larger vertical extent than the millimeter-continuum due to the larger millimeter sized dust grains being settled toward the mid-plane \citep[e.g.][]{zhang2017, Dutrey2017}. Equation~\ref{eq:I_scatsame} can be modified to assume that the line and continuum generating layers are separated, with the line generating layer closer to the observer (Fig.~\ref{fig:dust_line_iso}, right). This could be the case for disk surface gas that is lofted by hydrostatic equilibrium to greater heights than the settled dust-rich midplane.
When considering the effects of scattering the line emission will appear to be stronger than in the fully mixed case. In this two layer case, using $\chi_\mathrm{line}$ = 1, and again using $\tau_\mathrm{dust} \gg 1$, we can write:
\begin{equation}
\label{eq:I_scatlayer}
    \begin{aligned}
       I_\mathrm{L-C} = &\chi_\mathrm{line}B_\nu\left(T_\mathrm{ex}\right) \left( 1 - e^{-\tau_\mathrm{line}}\right) + \\
       &\chi_\mathrm{dust}B_\nu\left(T_\mathrm{dust}\right)e^{-\tau_\mathrm{line}} - \\
        &\chi_\mathrm{dust}B_\nu\left(T_\mathrm{dust}\right) \\
       = &B_\nu\left(T_\mathrm{iso}\right) \left( 1 - e^{-\tau_\mathrm{line}}\right) -\\ &\chi_\mathrm{dust}B_\nu\left(T_\mathrm{dust}\right)\left( 1 - e^{-\tau_\mathrm{line}}\right)\\
       = &B_\nu\left(T_\mathrm{iso}\right) \left( 1 - e^{-\tau_\mathrm{line}}\right)\left(1 - \chi_\mathrm{dust}\right),
\end{aligned}
\end{equation}
In this case, there thus will always be line emission visible above an optically thick dust continuum unless the dust does not scatter its own radiation, that is, $\chi_\mathrm{dust} \approx 1$ \citep[which reduces to the case discussed in][]{Rosotti2021}.

\subsection{Fully suppressing line emission with optically thick dust}
\label{sss:linesuppression}

What is left to understand is under which physical conditions it is possible to suppress the line flux significantly. In both Eqs.~\ref{eq:I_scatsame}~and~\ref{eq:I_scatlayer} the line flux drops significantly when $\chi_\mathrm{dust}$ approaches 1, which implies negligible scattering. For particles of astronomically relevant composition and assuming $p = -3.5$, this happens when the maximum grain size is smaller than 0.1 mm, or larger than 10 cm \citep{Zhu2019}. Grains of these sizes, however, have at least an order of magnitude lower total (absorption+scattering) opacity compared to millimeter sized grains, implying that 10$\times$ more mass is needed to make the mid-plane optically thick at 1-3 millimeter wavelengths \citep[][]{sierra20}. This would imply large pileups of dust in the regions that are optically thick. 

Line flux could also be significantly reduced if the line emitting layer and continuum emitting layer are mixed. Here dust scattering and absorption  modifies the line flux. This would entail the lofting of millimeter sized grains to the layer from which the molecule of interest is emitting. In this case, dust scattering and absorption dominate the total opacity, even at line center. Taking the CO $J$=2--1 line as an example, we can estimate the local gas-to-dust ratios necessary. For simplicity a temperature of 50 K, CO line width of 1 km s$^{-1}$, and abundance with respect to \ce{H2} of $10^{-4}$ is assumed. Under these conditions CO gas has a gas-mass opacity of $\sim$1000 cm$^{2}$~g$^{-1}$ \citep[][]{Schoier2005, Endres2016}, compared to the 0.3 cm$^{2}$~g$^{-1}$ for millimeter sized dust at a gas-to-dust ratio of 100 \citep{Birnstiel2018}. \ce{^{12}CO} $J$=2--1 should thus only be affected by dust opacity when the gas-to-dust ratio drops to values around 0.03, which is unlikely. However, for \ce{^{13}CO} and \ce{C^{18}O} which are 69 and 550 times less abundant \citep{Wilson1994}, gas-to-dust ratios of $\sim$2 and $\sim$15 can cause significant suppression of the line flux. This assumes that the CO abundance is $10^{-4}$, for lower CO abundances, less dust would be necessary.

\section{Discussion}

\begin{figure}
    \centering
    \includegraphics[width = \hsize]{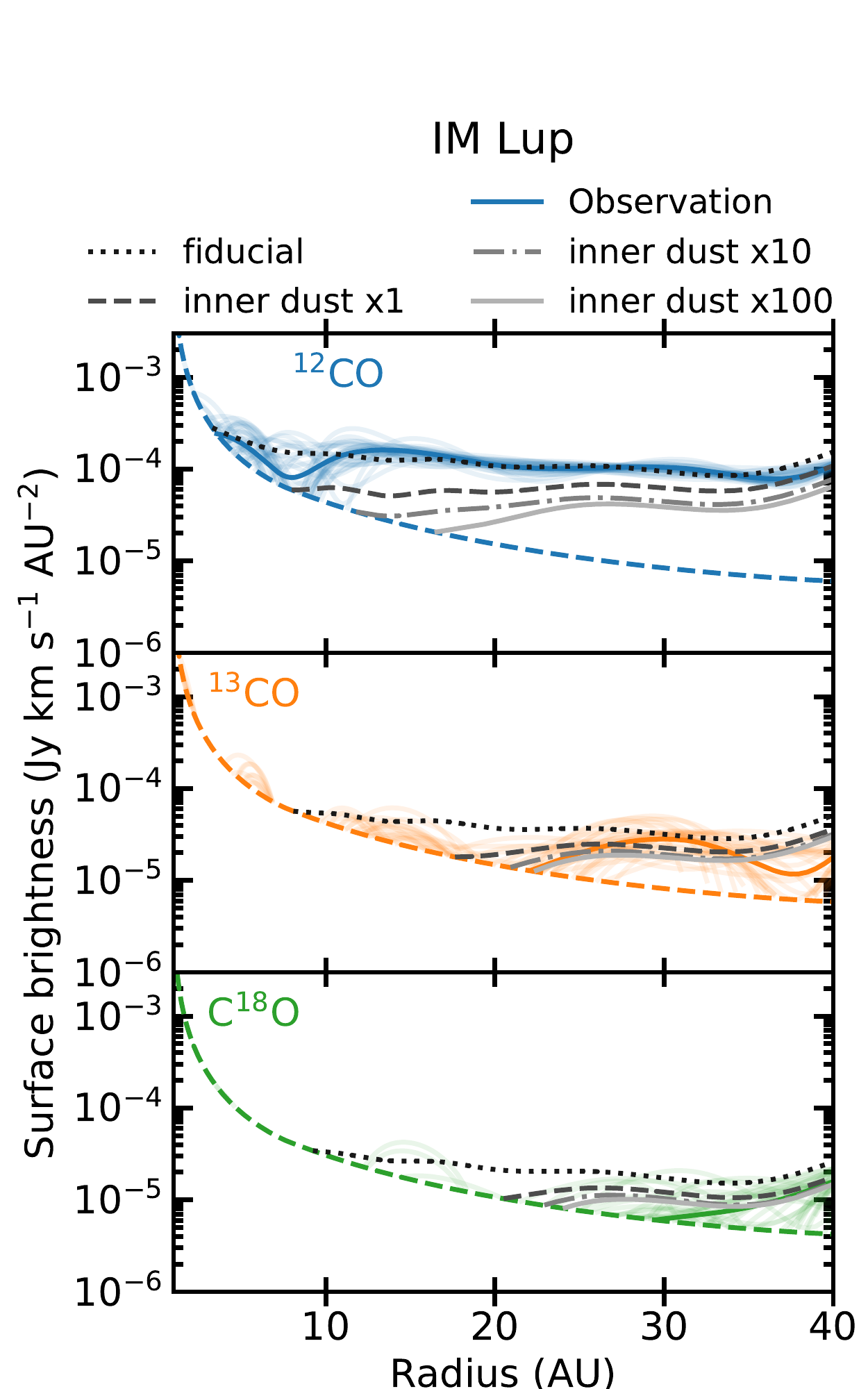}
    \caption{CO radial profiles of IM Lup compared with model radial profiles of \ce{^{12}CO} (top), \ce{^{13}CO} (middle) and \ce{C^{18}O} (bottom), for which the large dust vertical distribution is varied within 20 au. Colored solid lines show the CO isotopologue radial surface brightness profiles, and the dashed lines show the detection limit (see Fig.~\ref{fig:Radial_profiles}). Models are shown in grey scale in different line styles. Apart from the fiducial model, all models have the large dust fully vertically mixed with the gas, and the models shown with dashed-dotted and solid lines have the large dust within 20 au enhanced by a factor of 10 and 100 respectively. Surface densities and model description is given in Appendix~\ref{app:Toymodels}. }
    \label{fig:Inner_models}
\end{figure}

\subsection{Evidence for pebble drift feeding the inner IM Lup disk}
\label{ssc:IMLupinner}
IM Lup shows a strong surface brightness drop in \ce{^{13}CO} and \ce{C^{18}O} in the inner $\sim$30 au in the image derived radial profiles and in surface brightness profiles derived via the line profiles.   Dust models suggest that the dust continuum emission is optically thick in the inner regions \citep{Cleeves2017, sierra20}, with optical depths $>3$ within 40 au \citep{sierra20}. As argued in Sec.~\ref{sss:linesuppression}, for this dust to significantly affect the \ce{C^{18}O} emission, the gas-to-dust ratio must be lower than 100. In the case that the dust is non-scattering, at least 500 M$_\oplus$ is necessary to make the dust optically thick and thus suppress the line emission within the inner 20 au. The amount of dust necessary if the grains are around millimeter-sized and vertically extended is not straight forward to estimate due to the complex interaction between line emission and the scattering of the dust. As calculated in Sec.~\ref{sss:linesuppression}, 500 $M_\oplus$ which implies a gas-to-dust ratio of $\sim$10 is of the right magnitude to impact the line emission.  

To better constrain the mass necessary to reproduce the observations in the case of a vertically extended grain population we calculated the CO line emission in a couple of Dust And LInes (\rm{DALI}) models \citep[][]{Bruderer2012, Bruderer2013}. These toy-models are tuned to roughly match the CO column derived at $\sim$ 100 au radii by \citet{zhang20}, assuming a CO abundance of $10^{-5}$ (model setup is presented in Appendix~\ref{app:Toymodels}). This model is just a representative model to see the effects of dust, and is not finely tuned to the IM Lup (inner)disk. The distribution and amount of large dust within 20 au is varied in these models. In the fiducial model, the large dust (up to 1 mm) is settled, in all other models, this large dust population is well mixed in with the gas. At the same time, the amount of large dust is increased by a factor of 1, 10 and 100, exploring gas-to-dust ratios of 100, 10 and 1. The radial profiles derived from model line profiles, and the IM Lup observations as comparison are shown in Figure~\ref{fig:Inner_models}.  The goal of these models is not to fit the data perfectly, but to see the effects of different dust distributions on the CO isotopologue emission.

The model \ce{^{12}CO} surface brightness drops uniformly when the amount of dust in the atmosphere is increased. This is caused by the extra dust in the atmosphere lowering the gas and dust temperature, due to decreased penetration of optical and UV photons. The \ce{^{13}CO} and \ce{C^{18}O} emission in the models are also impacted by the additional large dust. Outside of 20 au, a brightness drop like that for \ce{^{12}CO} is seen. Inward of 20 au there is an extra drop in models with 10 and 100 times more dust. This allows us to reproduce the order of magnitude drop in \ce{C^{18}O} and \ce{^13{CO}} intensities.

The amount of dust necessary to obscure the line photons in the model is between 400 and 4000 M$_\oplus$ within 20 au. This is in agreement with the amount of dust derived from multi-wavelength continuum fitting, which finds $\sim$500 M$_\oplus$ of grains within 20 au \citep{sierra20}. The minimal mass necessary to explain the CO observations is thus $\sim$ 500 $M_\oplus$ regardless of grain size. However, if the grains have a significant scattering opacity, they will need to be vertically well mixed. Grains of 1 millimeter have a Stokes number of $ \mathrm{St} = 4\times 10^{-3}$ at 20 au ($\Sigma_\mathrm{gas} \approx 100$). To efficiently mix these into the surface layers we need a turbulent $\alpha > \mathrm{St} = 4\times 10^{-3}$ \citep{Dubrulle1995}. This is significantly higher than the turbulence expected for an inner disk deadzone \citep[e.g.][]{Gammie1996}.

The masses for the millimeter grain models assume that the \ce{CO} abundance in the inner 20 au is $10^{-5}$. If it were increased to $10^{-4}$, this would imply 10 times more large dust as well as a gas-to-dust ratio of 0.1-1. The model thus needs more dust than the minimum amounts estimated from the analytical derivation (Sec.~\ref{sss:linesuppression}).

Given the young $\sim$1 Myr age of IM Lup \citep{Mawet2012}, it seems reasonable that the large agglomeration of dust in the inner disk is due to radial drift of the grains. Radial drift can explain both the strong agglomeration of dust within the 1 Myr timescale as well as the specific transport of dust, but not gas, leading to low gas-to-dust ratios. Radial dust transport rates of $\gtrsim 100$ M$_\oplus$ per Myr are often invoked to quickly build the cores of gas giants in pebble accretion models \citep[e.g.][]{Bitsch2019}, which appears possible in the IM Lup disk.

The low gas-to-dust ratios derived in the inner disk would also allow for the quick formation of planetesimals through the steaming instability.   The streaming instability is most efficient when the grain(s) Stokes number is close to 0.1 \citep[e.g.][]{Carrera2015}.  This corresponds to 3-10 cm for IM Lup at 10 au assuming the surface density from \citet{zhang20}. This is on the large end of the size distribution consistent with the dust continuum modeling \citep[1 cm, assuming $p = -2.5$][]{sierra20}, but also agrees with large grain models that are weakly scattering \citep{Zhu2019}. Grains down to Stokes numbers of $3\times 10^{-4}$ ($\sim 0.1$ mm) have also been found to be able to trigger the streaming instability at a gas-to-dust ratio of 10, as measured for IM Lup.  Even if  grains are small in the inner disk triggering the streaming instability remains likely. Planetesimals could thus be currently forming through the streaming instability in the inner 20 au of IM Lup.

Strong pebble drift can have a profound impact on the composition of the gas in the inner disk \citep[e.g.][]{Ciesla2006, Booth2019}, with the pebbles bringing in oxygen-rich ices that lower the C/O ratio within the \ce{CO2} and \ce{H2O} ice lines. It has been proposed that these effects can be seen in mid-infrared spectra of these sources, by enhancement of the oxygen carrying species, \ce{H2O} and \ce{CO2}, and subsequent suppression of the carbon-rich species \citep[][]{Najita2013, Bosman2018,Banzatti2020}.

\begin{figure*}
    \centering
    \includegraphics[width = \hsize]{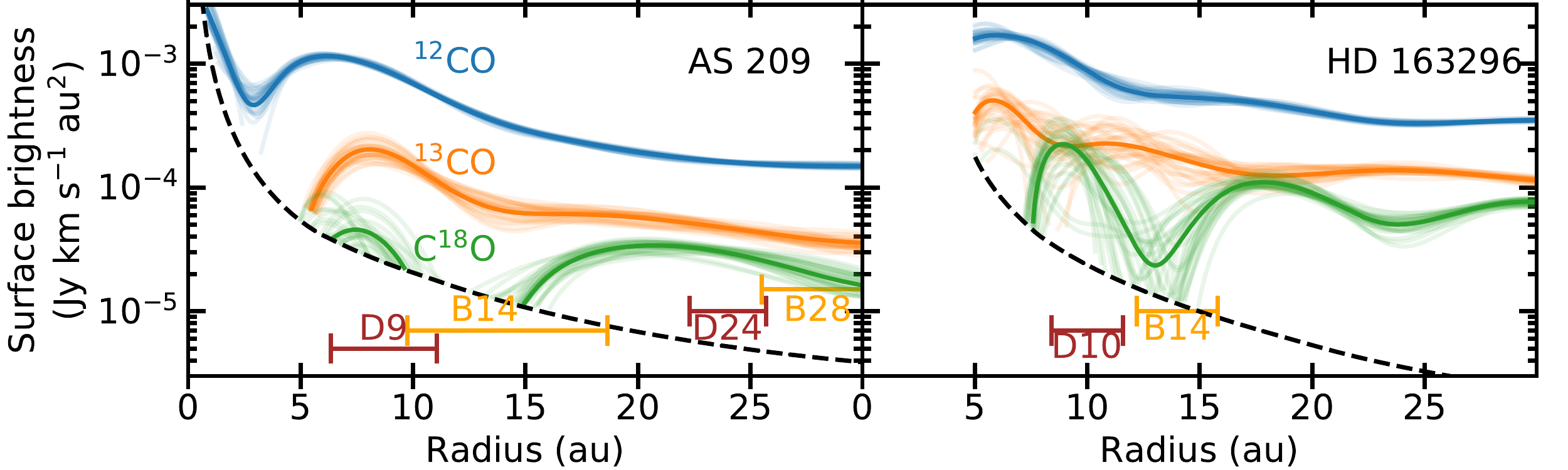}
    \caption{Line profile derived radial structures for AS 209 and HD 163296 together with dust substructures found in the DSHARP ALMA large program \citep{Andrews2018, Huang2018}. Locations and width of bright rings are given in orange, dark gaps in brown. Line colors and styles for the radial profiles are the same as in Fig.~\ref{fig:Radial_profiles}. In both AS 209 and HD 163296 \ce{C^{18}O} line emission peaks at the location of the dust gap while line emission is suppressed at the location of the dust ring.}
    \label{fig:AS_HD_structure}
\end{figure*}

The \textit{Spitzer}-IRS IM Lup spectrum shows a clear \ce{CO2} emission feature at 15 micron, but does not show any \ce{H2O}, \ce{OH}, \ce{C2H2}, or \ce{HCN} emission, making it a peculiar source in the T Tauri sample \citep{Pontoppidan2010, Salyk2011, Bosman2017}. It is not consistent with disks with strong water emission and weak HCN emission, as expected for a disk with strong drift \citep{Najita2013, Banzatti2020}. However, in the case of strong drift and vertical mixing, \citet{Bosman2018} found extremely elevated \ce{CO2} abundances, resulting in \ce{CO2}-branch fluxes far brighter than observed towards any disk.

The low gas-to-dust ratio inferred for the inner disk of IM Lup might solve this disparity. 
The strong line emission observed in many other T-Tauri disks in the mid-infrared which is originating from the surface layers of the inner few au, generally requires high ($>$ 1000) gas-to-dust ratios to reproduce \citep{Meijerink2009, Blevins2016, Bosman2017}. This is in sharp contrast to the gas-to-dust ratios of 0.1-10 derived from the ALMA data. If these low gas-to-dust ratios persist into the inner disk upper atmosphere, they would lower the \ce{CO2} flux. This would also smother the \ce{C2H2} and \ce{HCN} flux and could make the \ce{H2O} lines too weak to detect, leaving only the strong \ce{CO2} Q-branch. Sensitive JWST-MIRI spectra will be able to look for the weak \ce{H2O} lines to constrain the \ce{CO2} to \ce{H2O} abundance ratio.  These observations can potentially also detect \ce{H2} pure rotational lines to determine the absolute \ce{CO2} abundance and gas-to-dust ratio in the surface layers.

\subsection{Tracing the gas cavity in GM Aur}
In the millimeter continuum images the GM Aur disk stands out in the MAPS sample as it has a large cavity in the dust, with very little large dust within $\sim$40 au \citep[e.g.][]{Dutrey2008, Huang2020, sierra20}. This implies that there is not enough dust to affect the CO line emission. The surface brightness profiles show that the lines still emit from part of the dust cavity, but that there is a strong drop in surface brightness, at least a factor 3, in all isotopologues. This is significantly larger than the $\sim$30\% drop seen in the radial profiles extracted from the images (see Fig.~\ref{fig:Radial_profiles_w_image}).

Such strong emission drops can only be from column density drops strong enough to make the \ce{^{12}CO} and \ce{^{13}CO} lines optically thin. Assuming a \ce{^{12}CO} to \ce{^{13}CO} ratio of 69 \citep{Wilson1994}, this means that the CO column must drop by almost 2 orders of magnitude between 20 au, where the drop in \ce{^{13}CO} begins, and 15 au, where the drop in \ce{^{12}CO} begins. As CO strongly self-shields, it is likely that this is due to a total column density drop, with the \ce{H2} column dropping below $\sim 10^{21}$ cm$^{-2}$ around 15 au \citep{Bruderer2012, Bosman2019}, but source specific models, including isotope-specific photo-dissociation are necessary to extract a more detailed \ce{CO} and \ce{H2} column density profile.

\subsection{Impact of gaps and rings}
\label{ssc:AS209gap}

The \ce{C^{18}O} surface brightness profiles in the inner disks of both HD 163296 and AS 209 show gaps and peaks that line up with known dust substructures for these disks imaged at $\sim$ 40 mas \citep[see Fig.~\ref{fig:AS_HD_structure}][]{Andrews2018, Huang2018}. The bright continuum rings, which exist at 14 au for both disks line up with a strong, $>3\times$, decrease in the \ce{C^{18}O} surface brightness. In contrast, at the location of millimeter continuum gaps, at 9 and 10 au for AS 209 and HD 163296, respectively, there is an increase in \ce{C^{18}O} surface brightness. In AS 209, the 9 au millimeter continuum gap also corresponds to a peak of the \ce{^{12}CO} and \ce{^{13}CO} surface brightness. This correspondence between continuum rings, and line minima implies that the dust is impacting the line emission in these regions. This, in turn, implies that either the dust in these rings has very little scattering opacity at 1.3 mm (grain sizes $< 0.1$ mm or $> 10$ cm) or millimeter dust that is vertically extended (see Sec.~\ref{sss:linesuppression}). In both cases, the ring would need to contain a significantly larger surface density of dust, by at least factor 10, when compared to the surrounding material using the surface densities given by \citet{zhang20}. The presence of a ring signals efficient radial concentration of dust, vertical concentration of dust is also expected.  This argues against vertically extended efficiently scattering grains. Furthermore, small grains $<0.1$ mm would have small stokes numbers, and would thus not be efficiently trapped within the ring. As such efficient localized growth in the ring to multiple cm-sized grains seems to most logical explanation. 

The increase in \ce{C^{18}O} flux at the millimeter continuum gaps in AS 209 and HD 163296, is consistent with optically thin dust at these radii, with higher fluxes caused by the lower continuum flux.  Reduced continuum opacity allows for a higher column of CO to be probed. 
At the same time, it implies that the dust gap is rich in gas as a strongly gas depleted region at the location of the millimeter continuum gap would cause weak \ce{C^{18}O} emission from the gap. The peak of \ce{^{12}CO} and \ce{^{13}CO} in AS 209 around the 9 au millimeter continuum gap suggest that the gap does impact the temperature structure, with higher temperatures in and around this gap.
AS 209 also shows a strong drop in \ce{^{12}CO} surface density at $\sim$3 au where the \ce{^{13}CO} has already dropped under the detection threshold. The drop from the peak at 8 au is greater than a factor 2, implying either a similar drop in temperature or a drop of the CO column density below $\sim 10^{17}$ cm$^{-2}$, making CO optically thin. Interestingly, for the second continuum gap-ring pair at 24 and 28 au a clear effect on the \ce{C^{18}O} is missing.

\subsection{Unresolved dust structure in MWC 480?}
The MWC 480 disk shows a drop in \ce{C^{18}O} surface brightness and a peak in \ce{^{13}CO} surface brightness at 10 au. High resolution continuum observations do not show any structure in the inner regions of the MWC 480 disk \citep{Long2018, sierra20}. However, the highest resolution available is 0.08 mas, or 13 au, which would smear out any structure around 10 au.

The combination of a \ce{C^{18}O} drop and a \ce{^{13}CO} peak is difficult to explain. The increase in \ce{^{13}CO}, and to a lesser extent \ce{^{12}CO} implies an increase in gas temperature. The drop in \ce{C^{18}O} would imply an optically thin \ce{C^{18}O} column, while \ce{^{13}CO} is still optically thick. This would constrain the CO column density between 3$\times10^{18}$ and $2\times 10^{19}$~cm$^{-2}$. (assuming 50 K for the kinetic temperature). This is at least 3-4 orders of magnitude lower compared to the CO column derived by reproducing the CO isotopologue emission with a thermo-chemical model, and the column necessary to explain the hyperfine line-ratios of the \ce{C^{17}O} $J=$1-0 emission originating from the inner regions of the MWC 480 disk \citep{zhang20}. As such I might be more likely that we are tracing some more complex interplay between unresolved dust sub-structure and temperature.  

\subsection{Future directions}

Analysis of the dust in the inner regions of IM Lup, HD 163296 and MWC 480 in \citet{sierra20} suggests that the grains responsible for the millimeter emission might be small (0.1 mm).  These small grains have little scattering opacity at 1.3 and 3 mm and thus efficiently block the line emission (see Sec.~\ref{sss:linesuppression}). At higher frequency, the scattering opacity of these small grains increases relative to the absorption opacity. This would, counter intuitively, increase the expected line flux of high frequency lines, and would allow the use of \ce{^{13}CO} and \ce{C^{18}O} $J=$8-7 lines to constrain the grain size in the inner disk. At high frequency (877 GHz; ALMA, Band 10) integrations would be shorter, compared to observations with a similar sensitivity at lower frequency (110 GHz; ALMA Band 3).


Emission originating from the inner disk to needs to be bright to allow for the extraction high resolution surface brightness radial profiles of the inner $\sim$20 au from the spectra. This unfortunately constrains the available species that can be used. CO isotopologues, as used here, are the prime candidates. In disks with abundant dust in the inner 10s of au, HCN is the only other species bright enough to extract high resolution radial surface brightness profiles but it has the issue of satellite lines that complicate the analysis. In GM Aur, more species are centrally peaked, including CN, CS and \ce{C2H}.  This allows for a high resolution chemical study of the GM Aur disk cavity, and possibly other transition disks, without going to $\ll$ 0\farcs15 resolution. Lines that are weak in the inner disk it is still possible to get upper limits to the surface brightness profile, which can still provide valuable information \citep[see][]{bosman20_CtoO}.

Finally, using a space or moon based submillimeter telescope with a high resolving power instrument ($R = \lambda/\delta\lambda>10^6$) such as the Origins Space Telescope equipped with HERO (Heterodyne Receiver for Origins), it would be possible to use this technique on bright \ce{H2^{16}O} water lines, mapping water to far within the telescopes nominal spatial resolution.

\section{Conclusions}

We have studied the inner 20 au, using wings of the CO isotopologue lines observed with the MAPS program \citep{oberg20}. Using Keplerian rotation of the disk we have inferred radial surface brightness profiles with an effective resolution of $\sim$3 au. Our conclusions from these radial surface brightness profiles are as follows:
\begin{itemize}
    \item The $J$=1--0 and $J$=2--1 \ce{^{13}CO} and \ce{C^{18}O} line wings are consistent within the observational errors for all sources. This implies that the lower frequency $J$=1--0 lines do not probe significantly deeper down toward the midplane than the $J$=2--1 lines.  
    
    
    \item Radial surface brightness profiles extracted from the line profiles show a host of features, including gaps and peaks in the \ce{C^{18}O} surface brightness in AS 209, HD 163296 and MWC 480, an inner hole in \ce{^{13}CO} and \ce{C^{18}O} in IM Lup (20-30 au) and AS 209 (5 au), as well as an inner hole in \ce{^{12}CO}, \ce{^{13}CO} and \ce{C^{18}O} in GM Aur. Some of these features are not resolved in the CLEANED images, showcasing the power of this technique.
    
    \item The central depletion of CO isotopologue emission seen for GM Aur is consistent with a gas cavity within the dust cavity as previously inferred \citep[e.g.][]{Dutrey2008}. We estimate that \ce{^{12}CO} is optically thin within 15 au.
    
    \item The inner hole in the CO emission in IM Lup can be explained by a pileup of large dust in the inner 20 au. This is in line with multi-wavelength continuum emission analysis and implies that there is $> 400$ $M_\oplus$ of large dust and a gas-to-dust ratio of $\lesssim$10 within 20 au in the IM Lup disk. We propose that this pileup is due to radial drift of large dust. The drift rate necessary to to cause this pileup, $> 100\,M_\oplus \mathrm{Myr}^{-1}$, enables the quick formation of giant planet cores through pebble accretion, while the small gas-to-dust ratios would allow for the triggering of the streaming instability.
    
    \item The previously resolved continuum rings in the inner 20 au of AS 209 and HD 163296 appear to impact the \ce{CO} isotopologue emission. The innermost bright continuum ring in both systems is co-spatial with a drop in the \ce{C^{18}O} surface brightness, while the millimeter continuum gap inward of this show an emission peak in all isotopologues in both disks. Efficient growth of grains to multiple cm sizes within the continuum ring, coupled with trapping of these grains can explain the lack of CO emission.
    
    \item The CO isotopologue surface brightness in the inner disk of MWC 480 shows a strong variation in the \ce{^{13}CO} and \ce{C^{18}O} surface brightness at 10 au for which no explanation has been found.

\end{itemize}

\acknowledgments

This paper makes use of the following ALMA data: ADS/JAO.ALMA\#2018.1.01055.L. ALMA is a partnership of ESO (representing its member states), NSF (USA) and NINS (Japan), together with NRC (Canada), MOST and ASIAA (Taiwan), and KASI (Republic of Korea), in cooperation with the Republic of Chile. The Joint ALMA Observatory is operated by ESO, AUI/NRAO and NAOJ. The National Radio Astronomy Observatory is a facility of the National Science Foundation operated under cooperative agreement by Associated Universities, Inc.

A.D.B., E.A.B., F.A., and K.I.Ö. acknowledge support from NSF AAG grant No. 1907653.
S.M.A. and J.H. acknowledge funding support from the National Aeronautics and Space Administration under Grant No. 17-XRP17 2-0012 issued through the Exoplanets Research Program.
M.L.R.H. acknowledges support from the Michigan Society of Fellows
R.T. \& F.L. acknowledge support from the Smithsonian Institution as a Submillimeter Array (SMA) Fellow.
K.I.\"O. acknowledges support from the Simons Foundation (SCOL No. 321183).
V.V.G. acknowledges support from FONDECYT Iniciaci\'on 11180904 and ANID project Basal AFB-170002.
C.W.~acknowledges financial support from the University of Leeds and from the Science and Technology Facilities Council (grant numbers ST/R000549/1 and ST/T000287/1).
Y.A. acknowledges support by NAOJ ALMA Scientific Research Grant Code 2019-13B, and Grant-in-Aid for Scientific Research 18H05222 and 20H05847.
J.B., J.B.B., I.C., J.H., K.R.S. \& K.Z. acknowledge support by NASA through the NASA Hubble Fellowship grants \#HST-HF2-51427.001-A, \#HST-HF2-51429.001-A, HST-HF2-51405.001-A, \#HST-HF2-51460.001-A, \#HST-HF2-51419.001 \& HST-HF2-51401.001. awarded by  the  Space  Telescope  Science  Institute,  which  is  operated  by  the  Association  of  Universities  for  Research  in  Astronomy, Incorporated, under NASA contract NAS5-26555.
A.S.B acknowledges the studentship funded by the Science and Technology Facilities Council of the United Kingdom (STFC).

G.C. is supported by NAOJ ALMA Scientific Research Grant Code. 2019-13B. 
L.I.C. gratefully acknowledges support from the David and Lucille Packard Foundation and Johnson \& Johnson's WiSTEM2D Program.
J.D.I. acknowledges support from the Science and Technology Facilities Council of the United Kingdom (STFC) under ST/T000287/1.
C.J.L. acknowledges funding from the National Science Foundation Graduate Research Fellowship under Grant No. DGE1745303.
Y.L. acknowledges the financial support by the Natural Science Foundation of China (Grant No. 11973090).
F.M. acknowledges support from ANR of France under contract ANR-16-CE31-0013 (Planet-Forming-Disks)  and ANR-15-IDEX-02 (through CDP ``Origins of Life'').
H.N. acknowledges support by NAOJ ALMA Scientific Research Grant Code 2018-10B and Grant-in-Aid for Scientific Research 18H05441.
L.M.P.\ acknowledges support from ANID project Basal AFB-170002 and from ANID FONDECYT Iniciaci\'on project \#11181068.
A.S. acknowledges support from ANID/CONICYT Programa de Astronom\'ia Fondo ALMA-CONICYT 2018 31180052.
T.T. is supported by JSPS KAKENHI Grant Numbers JP17K14244 and JP20K04017.
Y.Y. is supported by IGPEES, WINGS Program, the University of Tokyo.
K.Z. acknowledges the support of the Office of the Vice Chancellor for Research and Graduate Education at the University of Wisconsin – Madison with funding from the Wisconsin Alumni Research Foundation. 

\bibliographystyle{aasjournal}
\bibliography{Lit_list}

\begin{thebibliography}{}
\expandafter\ifx\csname natexlab\endcsname\relax\def\natexlab#1{#1}\fi
\providecommand{\url}[1]{\href{#1}{#1}}
\providecommand{\dodoi}[1]{doi:~\href{http://doi.org/#1}{\nolinkurl{#1}}}
\providecommand{\doeprint}[1]{\href{http://ascl.net/#1}{\nolinkurl{http://ascl.net/#1}}}
\providecommand{\doarXiv}[1]{\href{https://arxiv.org/abs/#1}{\nolinkurl{https://arxiv.org/abs/#1}}}

\bibitem[{{ALMA Partnership} {et~al.}(2015){ALMA Partnership}, {Brogan},
  {P{\'e}rez}, {Hunter}, {Dent}, {Hales}, {Hills}, {Corder}, {Fomalont},
  {Vlahakis}, {Asaki}, {Barkats}, {Hirota}, {Hodge}, {Impellizzeri}, {Kneissl},
  {Liuzzo}, {Lucas}, {Marcelino}, {Matsushita}, {Nakanishi}, {Phillips},
  {Richards}, {Toledo}, {Aladro}, {Broguiere}, {Cortes}, {Cortes}, {Espada},
  {Galarza}, {Garcia-Appadoo}, {Guzman-Ramirez}, {Humphreys}, {Jung}, {Kameno},
  {Laing}, {Leon}, {Marconi}, {Mignano}, {Nikolic}, {Nyman}, {Radiszcz},
  {Remijan}, {Rod{\'o}n}, {Sawada}, {Takahashi}, {Tilanus}, {Vila Vilaro},
  {Watson}, {Wiklind}, {Akiyama}, {Chapillon}, {de Gregorio-Monsalvo}, {Di
  Francesco}, {Gueth}, {Kawamura}, {Lee}, {Nguyen Luong}, {Mangum}, {Pietu},
  {Sanhueza}, {Saigo}, {Takakuwa}, {Ubach}, {van Kempen}, {Wootten},
  {Castro-Carrizo}, {Francke}, {Gallardo}, {Garcia}, {Gonzalez}, {Hill},
  {Kaminski}, {Kurono}, {Liu}, {Lopez}, {Morales}, {Plarre}, {Schieven},
  {Testi}, {Videla}, {Villard}, {Andreani}, {Hibbard}, \&
  {Tatematsu}}]{ALMA2015}
{ALMA Partnership}, {Brogan}, C.~L., {P{\'e}rez}, L.~M., {et~al.} 2015, \apj,
  808, L3, \dodoi{10.1088/2041-8205/808/1/L3}

\bibitem[{{Andrews} {et~al.}(2018){Andrews}, {Huang}, {P{\'e}rez}, {Isella},
  {Dullemond}, {Kurtovic}, {Guzm{\'a}n}, {Carpenter}, {Wilner}, {Zhang}, {Zhu},
  {Birnstiel}, {Bai}, {Benisty}, {Hughes}, {{\"O}berg}, \&
  {Ricci}}]{Andrews2018}
{Andrews}, S.~M., {Huang}, J., {P{\'e}rez}, L.~M., {et~al.} 2018, \apj, 869,
  L41, \dodoi{10.3847/2041-8213/aaf741}

\bibitem[{{Armitage}(2011)}]{Armitage2011}
{Armitage}, P.~J. 2011, \araa, 49, 195,
  \dodoi{10.1146/annurev-astro-081710-102521}

\bibitem[{{Avenhaus} {et~al.}(2017){Avenhaus}, {Quanz}, {Schmid}, {Dominik},
  {Stolker}, {Ginski}, {de Boer}, {Szul{\'a}gyi}, {Garufi}, {Zurlo},
  {Hagelberg}, {Benisty}, {Henning}, {M{\'e}nard}, {Meyer}, {Baruffolo},
  {Bazzon}, {Beuzit}, {Costille}, {Dohlen}, {Girard}, {Gisler}, {Kasper},
  {Mouillet}, {Pragt}, {Roelfsema}, {Salasnich}, \& {Sauvage}}]{Avenhaus2017}
{Avenhaus}, H., {Quanz}, S.~P., {Schmid}, H.~M., {et~al.} 2017, \aj, 154, 33,
  \dodoi{10.3847/1538-3881/aa7560}

\bibitem[{{Banzatti} \& {Pontoppidan}(2015)}]{Banzatti2015}
{Banzatti}, A., \& {Pontoppidan}, K.~M. 2015, \apj, 809, 167,
  \dodoi{10.1088/0004-637X/809/2/167}

\bibitem[{{Banzatti} {et~al.}(2017){Banzatti}, {Pontoppidan}, {Salyk},
  {Herczeg}, {van Dishoeck}, \& {Blake}}]{Banzatti2017}
{Banzatti}, A., {Pontoppidan}, K.~M., {Salyk}, C., {et~al.} 2017, \apj, 834,
  152, \dodoi{10.3847/1538-4357/834/2/152}

\bibitem[{{Banzatti} {et~al.}(2020){Banzatti}, {Pascucci}, {Bosman}, {Pinilla},
  {Salyk}, {Herczeg}, {Pontoppidan}, {Vazquez}, {Watkins}, {Krijt}, {Hendler},
  \& {Long}}]{Banzatti2020}
{Banzatti}, A., {Pascucci}, I., {Bosman}, A.~D., {et~al.} 2020, \apj, 903, 124,
  \dodoi{10.3847/1538-4357/abbc1a}

\bibitem[{{Birnstiel} {et~al.}(2018){Birnstiel}, {Dullemond}, {Zhu}, {Andrews},
  {Bai}, {Wilner}, {Carpenter}, {Huang}, {Isella}, {Benisty}, {P{\'e}rez}, \&
  {Zhang}}]{Birnstiel2018}
{Birnstiel}, T., {Dullemond}, C.~P., {Zhu}, Z., {et~al.} 2018, \apjl, 869, L45,
  \dodoi{10.3847/2041-8213/aaf743}

\bibitem[{{Bitsch} {et~al.}(2019){Bitsch}, {Izidoro}, {Johansen}, {Raymond},
  {Morbidelli}, {Lambrechts}, \& {Jacobson}}]{Bitsch2019}
{Bitsch}, B., {Izidoro}, A., {Johansen}, A., {et~al.} 2019, \aap, 623, A88,
  \dodoi{10.1051/0004-6361/201834489}

\bibitem[{{Blevins} {et~al.}(2016){Blevins}, {Pontoppidan}, {Banzatti},
  {Zhang}, {Najita}, {Carr}, {Salyk}, \& {Blake}}]{Blevins2016}
{Blevins}, S.~M., {Pontoppidan}, K.~M., {Banzatti}, A., {et~al.} 2016, \apj,
  818, 22, \dodoi{10.3847/0004-637X/818/1/22}

\bibitem[{{Booth} \& {Ilee}(2019)}]{Booth2019}
{Booth}, R.~A., \& {Ilee}, J.~D. 2019, \mnras, 487, 3998,
  \dodoi{10.1093/mnras/stz1488}

\bibitem[{{Bosman} {et~al.}(2019){Bosman}, {Banzatti}, {Bruderer}, {Tielens},
  {Blake}, \& {van Dishoeck}}]{Bosman2019}
{Bosman}, A.~D., {Banzatti}, A., {Bruderer}, S., {et~al.} 2019, \aap, 631,
  A133, \dodoi{10.1051/0004-6361/201935910}

\bibitem[{{Bosman} {et~al.}(2017){Bosman}, {Bruderer}, \& {van
  Dishoeck}}]{Bosman2017}
{Bosman}, A.~D., {Bruderer}, S., \& {van Dishoeck}, E.~F. 2017, A\&A, 601, A36,
  \dodoi{10.1051/0004-6361/201629946}

\bibitem[{{Bosman} {et~al.}(2018){Bosman}, {Tielens}, \& {van
  Dishoeck}}]{Bosman2018}
{Bosman}, A.~D., {Tielens}, A.~G.~G.~M., \& {van Dishoeck}, E.~F. 2018, \aap,
  611, A80, \dodoi{10.1051/0004-6361/201732056}

\bibitem[{{Bosman} {et~al.}(2021){Bosman}, {Alarc{\'o}n}, {Bergin}, {Zhang},
  {van 't Hoff}, {{\"O}berg}, {Guzm{\'a}n}, {Walsh}, {Aikawa}, {Andrews},
  {Bergner}, {Booth}, {Cataldi}, {Cleeves}, {Czekala}, {Furuya}, {Huang},
  {Ilee}, {Law}, {Le Gal}, {Liu}, {Long}, {Loomis}, {M{\'e}nard}, {Nomura},
  {Qi}, {Schwarz}, {Teague}, {Tsukagoshi}, {Yamato}, \&
  {Wilner}}]{bosman20_CtoO}
{Bosman}, A.~D., {Alarc{\'o}n}, F., {Bergin}, E.~A., {et~al.} 2021, arXiv
  e-prints, arXiv:2109.06221.
\newblock \doarXiv{2109.06221}

\bibitem[{{Brown} {et~al.}(2013){Brown}, {Pontoppidan}, {van Dishoeck},
  {Herczeg}, {Blake}, \& {Smette}}]{Brown2013}
{Brown}, J.~M., {Pontoppidan}, K.~M., {van Dishoeck}, E.~F., {et~al.} 2013,
  \apj, 770, 94, \dodoi{10.1088/0004-637X/770/2/94}

\bibitem[{{Bruderer}(2013)}]{Bruderer2013}
{Bruderer}, S. 2013, \aap, 559, A46, \dodoi{10.1051/0004-6361/201321171}

\bibitem[{{Bruderer} {et~al.}(2012){Bruderer}, {van Dishoeck}, {Doty}, \&
  {Herczeg}}]{Bruderer2012}
{Bruderer}, S., {van Dishoeck}, E.~F., {Doty}, S.~D., \& {Herczeg}, G.~J. 2012,
  \aap, 541, A91, \dodoi{10.1051/0004-6361/201118218}

\bibitem[{{Carmona} {et~al.}(2017){Carmona}, {Thi}, {Kamp}, {Baruteau},
  {Matter}, {van den Ancker}, {Pinte}, {K{\'o}sp{\'a}l}, {Audard}, {Liebhart},
  {Sicilia-Aguilar}, {Pinilla}, {Reg{\'a}ly}, {G{\"u}del}, {Henning}, {Cieza},
  {Baldovin-Saavedra}, {Meeus}, \& {Eiroa}}]{Carmona2017}
{Carmona}, A., {Thi}, W.~F., {Kamp}, I., {et~al.} 2017, \aap, 598, A118,
  \dodoi{10.1051/0004-6361/201628472}

\bibitem[{{Carr} \& {Najita}(2008)}]{Carr2008}
{Carr}, J.~S., \& {Najita}, J.~R. 2008, Science, 319, 1504,
  \dodoi{10.1126/science.1153807}

\bibitem[{{Carrera} {et~al.}(2015){Carrera}, {Johansen}, \&
  {Davies}}]{Carrera2015}
{Carrera}, D., {Johansen}, A., \& {Davies}, M.~B. 2015, \aap, 579, A43,
  \dodoi{10.1051/0004-6361/201425120}

\bibitem[{{Ciesla} \& {Cuzzi}(2006)}]{Ciesla2006}
{Ciesla}, F.~J., \& {Cuzzi}, J.~N. 2006, \icarus, 181, 178,
  \dodoi{10.1016/j.icarus.2005.11.009}

\bibitem[{{Cleeves} {et~al.}(2017){Cleeves}, {Bergin}, {{\"O}berg}, {Andrews},
  {Wilner}, \& {Loomis}}]{Cleeves2017}
{Cleeves}, L.~I., {Bergin}, E.~A., {{\"O}berg}, K.~I., {et~al.} 2017, \apjl,
  843, L3, \dodoi{10.3847/2041-8213/aa76e2}

\bibitem[{{Cridland} {et~al.}(2019){Cridland}, {Pudritz}, \&
  {Alessi}}]{Cridland2019}
{Cridland}, A.~J., {Pudritz}, R.~E., \& {Alessi}, M. 2019, \mnras, 484, 345,
  \dodoi{10.1093/mnras/stz008}

\bibitem[{{Czekala} {et~al.}(2021){Czekala}, {Loomis}, {Teague}, {Booth},
  {Huang}, {Cataldi}, {Ilee}, {Law}, {Walsh}, {Bosman}, {Guzm{\'a}n}, {Le Gal},
  {{\"O}berg}, {Yamato}, {Aikawa}, {Andrews}, {Bae}, {Bergin}, {Bergner},
  {Cleeves}, {Kurtovic}, {M{\'e}nard}, {Nomura}, {P{\'e}rez}, {Qi}, {Schwarz},
  {Tsukagoshi}, {Waggoner}, {Wilner}, \& {Zhang}}]{czekala20}
{Czekala}, I., {Loomis}, R.~A., {Teague}, R., {et~al.} 2021, arXiv e-prints,
  arXiv:2109.06188.
\newblock \doarXiv{2109.06188}

\bibitem[{{Dubrulle} {et~al.}(1995){Dubrulle}, {Morfill}, \&
  {Sterzik}}]{Dubrulle1995}
{Dubrulle}, B., {Morfill}, G., \& {Sterzik}, M. 1995, \icarus, 114, 237,
  \dodoi{10.1006/icar.1995.1058}

\bibitem[{{Dullemond} \& {Monnier}(2010)}]{Dullemond2010}
{Dullemond}, C.~P., \& {Monnier}, J.~D. 2010, \araa, 48, 205,
  \dodoi{10.1146/annurev-astro-081309-130932}

\bibitem[{{Dutrey} {et~al.}(2008){Dutrey}, {Guilloteau}, {Pi{\'e}tu},
  {Chapillon}, {Gueth}, {Henning}, {Launhardt}, {Pavlyuchenkov}, {Schreyer}, \&
  {Semenov}}]{Dutrey2008}
{Dutrey}, A., {Guilloteau}, S., {Pi{\'e}tu}, V., {et~al.} 2008, \aap, 490, L15,
  \dodoi{10.1051/0004-6361:200810732}

\bibitem[{{Dutrey} {et~al.}(2017){Dutrey}, {Guilloteau}, {Pi{\'e}tu},
  {Chapillon}, {Wakelam}, {Di Folco}, {Stoecklin}, {Denis-Alpizar}, {Gorti},
  {Teague}, {Henning}, {Semenov}, \& {Grosso}}]{Dutrey2017}
---. 2017, \aap, 607, A130, \dodoi{10.1051/0004-6361/201730645}

\bibitem[{{Endres} {et~al.}(2016){Endres}, {Schlemmer}, {Schilke}, {Stutzki},
  \& {M{\"u}ller}}]{Endres2016}
{Endres}, C.~P., {Schlemmer}, S., {Schilke}, P., {Stutzki}, J., \&
  {M{\"u}ller}, H. S.~P. 2016, Journal of Molecular Spectroscopy, 327, 95,
  \dodoi{10.1016/j.jms.2016.03.005}

\bibitem[{{Ercolano} \& {Pascucci}(2017)}]{Ercolano2017}
{Ercolano}, B., \& {Pascucci}, I. 2017, Royal Society Open Science, 4, 170114,
  \dodoi{10.1098/rsos.170114}

\bibitem[{{Fernandes} {et~al.}(2019){Fernandes}, {Mulders}, {Pascucci},
  {Mordasini}, \& {Emsenhuber}}]{Fernandes2019}
{Fernandes}, R.~B., {Mulders}, G.~D., {Pascucci}, I., {Mordasini}, C., \&
  {Emsenhuber}, A. 2019, \apj, 874, 81, \dodoi{10.3847/1538-4357/ab0300}

\bibitem[{{Gammie}(1996)}]{Gammie1996}
{Gammie}, C.~F. 1996, \apj, 457, 355, \dodoi{10.1086/176735}

\bibitem[{{Gravity Collaboration} {et~al.}(2017){Gravity Collaboration},
  {Abuter}, {Accardo}, {Amorim}, {Anugu}, {{\'A}vila}, {Azouaoui}, {Benisty},
  {Berger}, {Blind}, {Bonnet}, {Bourget}, {Brandner}, {Brast}, {Buron},
  {Burtscher}, {Cassaing}, {Chapron}, {Choquet}, {Cl{\'e}net}, {Collin},
  {Coud{\'e} Du Foresto}, {de Wit}, {de Zeeuw}, {Deen},
  {Delplancke-Str{\"o}bele}, {Dembet}, {Derie}, {Dexter}, {Duvert}, {Ebert},
  {Eckart}, {Eisenhauer}, {Esselborn}, {F{\'e}dou}, {Finger}, {Garcia}, {Garcia
  Dabo}, {Garcia Lopez}, {Gendron}, {Genzel}, {Gillessen}, {Gonte}, {Gordo},
  {Grould}, {Gr{\"o}zinger}, {Guieu}, {Haguenauer}, {Hans}, {Haubois}, {Haug},
  {Haussmann}, {Henning}, {Hippler}, {Horrobin}, {Huber}, {Hubert}, {Hubin},
  {Hummel}, {Jakob}, {Janssen}, {Jochum}, {Jocou}, {Kaufer}, {Kellner},
  {Kendrew}, {Kern}, {Kervella}, {Kiekebusch}, {Klein}, {Kok}, {Kolb}, {Kulas},
  {Lacour}, {Lapeyr{\`e}re}, {Lazareff}, {Le Bouquin}, {L{\`e}na}, {Lenzen},
  {L{\'e}v{\^e}que}, {Lippa}, {Magnard}, {Mehrgan}, {Mellein}, {M{\'e}rand},
  {Moreno-Ventas}, {Moulin}, {M{\"u}ller}, {M{\"u}ller}, {Neumann}, {Oberti},
  {Ott}, {Pallanca}, {Panduro}, {Pasquini}, {Paumard}, {Percheron}, {Perraut},
  {Perrin}, {Pfl{\"u}ger}, {Pfuhl}, {Phan Duc}, {Plewa}, {Popovic}, {Rabien},
  {Ram{\'{\i}}rez}, {Ramos}, {Rau}, {Riquelme}, {Rohloff}, {Rousset},
  {Sanchez-Bermudez}, {Scheithauer}, {Sch{\"o}ller}, {Schuhler}, {Spyromilio},
  {Straubmeier}, {Sturm}, {Suarez}, {Tristram}, {Ventura}, {Vincent},
  {Waisberg}, {Wank}, {Weber}, {Wieprecht}, {Wiest}, {Wiezorrek}, {Wittkowski},
  {Woillez}, {Wolff}, {Yazici}, {Ziegler}, \& {Zins}}]{Gravity2017}
{Gravity Collaboration}, {Abuter}, R., {Accardo}, M., {et~al.} 2017, \aap, 602,
  A94, \dodoi{10.1051/0004-6361/201730838}

\bibitem[{{Guzm{\'a}n} {et~al.}(2018){Guzm{\'a}n}, {Huang}, {Andrews},
  {Isella}, {P{\'e}rez}, {Carpenter}, {Dullemond}, {Ricci}, {Birnstiel},
  {Zhang}, {Zhu}, {Bai}, {Benisty}, {{\"O}berg}, \& {Wilner}}]{Guzman2018}
{Guzm{\'a}n}, V.~V., {Huang}, J., {Andrews}, S.~M., {et~al.} 2018, \apjl, 869,
  L48, \dodoi{10.3847/2041-8213/aaedae}

\bibitem[{{Hales} {et~al.}(2019){Hales}, {Gorti}, {Carpenter}, {Hughes}, \&
  {Flaherty}}]{Hales2019}
{Hales}, A.~S., {Gorti}, U., {Carpenter}, J.~M., {Hughes}, M., \& {Flaherty},
  K. 2019, \apj, 878, 113, \dodoi{10.3847/1538-4357/ab211e}

\bibitem[{{Huang} {et~al.}(2017){Huang}, {{\"O}berg}, {Qi}, {Aikawa},
  {Andrews}, {Furuya}, {Guzm{\'a}n}, {Loomis}, {van Dishoeck}, \&
  {Wilner}}]{Huang2017}
{Huang}, J., {{\"O}berg}, K.~I., {Qi}, C., {et~al.} 2017, \apj, 835, 231,
  \dodoi{10.3847/1538-4357/835/2/231}

\bibitem[{{Huang} {et~al.}(2018){Huang}, {Andrews}, {Dullemond}, {Isella},
  {P{\'e}rez}, {Guzm{\'a}n}, {{\"O}berg}, {Zhu}, {Zhang}, {Bai}, {Benisty},
  {Birnstiel}, {Carpenter}, {Hughes}, {Ricci}, {Weaver}, \&
  {Wilner}}]{Huang2018}
{Huang}, J., {Andrews}, S.~M., {Dullemond}, C.~P., {et~al.} 2018, \apjl, 869,
  L42, \dodoi{10.3847/2041-8213/aaf740}

\bibitem[{{Huang} {et~al.}(2020){Huang}, {Andrews}, {Dullemond}, {{\"O}berg},
  {Qi}, {Zhu}, {Birnstiel}, {Carpenter}, {Isella}, {Mac{\'\i}as}, {McClure},
  {P{\'e}rez}, {Teague}, {Wilner}, \& {Zhang}}]{Huang2020}
---. 2020, \apj, 891, 48, \dodoi{10.3847/1538-4357/ab711e}

\bibitem[{{Huang} {et~al.}(2021){Huang}, {Bergin}, {{\"O}berg}, {Andrews},
  {Teague}, {Law}, {Kalas}, {Aikawa}, {Bae}, {Bergner}, {Booth}, {Bosman},
  {Calahan}, {Cataldi}, {Cleeves}, {Czekala}, {Ilee}, {Le Gal}, {Guzm{\'a}n},
  {Long}, {Loomis}, {M{\'e}nard}, {Nomura}, {Qi}, {Schwarz}, {Tsukagoshi}, {van
  't Hoff}, {Walsh}, {Wilner}, {Yamato}, \& {Zhang}}]{huang20}
{Huang}, J., {Bergin}, E.~A., {{\"O}berg}, K.~I., {et~al.} 2021, arXiv
  e-prints, arXiv:2109.06224.
\newblock \doarXiv{2109.06224}

\bibitem[{{Johnson} {et~al.}(2010){Johnson}, {Aller}, {Howard}, \&
  {Crepp}}]{Johnson2010}
{Johnson}, J.~A., {Aller}, K.~M., {Howard}, A.~W., \& {Crepp}, J.~R. 2010,
  \pasp, 122, 905, \dodoi{10.1086/655775}

\bibitem[{{Klaassen} {et~al.}(2013){Klaassen}, {Juhasz}, {Mathews}, {Mottram},
  {De Gregorio-Monsalvo}, {van Dishoeck}, {Takahashi}, {Akiyama}, {Chapillon},
  {Espada}, {Hales}, {Hogerheijde}, {Rawlings}, {Schmalzl}, \&
  {Testi}}]{Klaassen2013}
{Klaassen}, P.~D., {Juhasz}, A., {Mathews}, G.~S., {et~al.} 2013, \aap, 555,
  A73, \dodoi{10.1051/0004-6361/201321129}

\bibitem[{{Law} {et~al.}(2021){Law}, {Loomis}, {Teague}, {{\"O}berg},
  {Czekala}, {Andrews}, {Huang}, {Aikawa}, {Alarc{\'o}n}, {Bae}, {Bergin},
  {Bergner}, {Boehler}, {Booth}, {Bosman}, {Calahan}, {Cataldi}, {Cleeves},
  {Furuya}, {Guzm{\'a}n}, {Ilee}, {Le Gal}, {Liu}, {Long}, {M{\'e}nard},
  {Nomura}, {Qi}, {Schwarz}, {Sierra}, {Tsukagoshi}, {Yamato}, {van't Hoff},
  {Walsh}, {Wilner}, \& {Zhang}}]{law20_rad}
{Law}, C.~J., {Loomis}, R.~A., {Teague}, R., {et~al.} 2021, arXiv e-prints,
  arXiv:2109.06210.
\newblock \doarXiv{2109.06210}

\bibitem[{{Lazareff} {et~al.}(2017){Lazareff}, {Berger}, {Kluska}, {Le
  Bouquin}, {Benisty}, {Malbet}, {Koen}, {Pinte}, {Thi}, {Absil}, {Baron},
  {Delboulb{\'e}}, {Duvert}, {Isella}, {Jocou}, {Juhasz}, {Kraus}, {Lachaume},
  {M{\'e}nard}, {Millan-Gabet}, {Monnier}, {Moulin}, {Perraut}, {Rochat},
  {Soulez}, {Tallon}, {Thi{\'e}baut}, {Traub}, \& {Zins}}]{Lazareff2017}
{Lazareff}, B., {Berger}, J.-P., {Kluska}, J., {et~al.} 2017, \aap, 599, A85,
  \dodoi{10.1051/0004-6361/201629305}

\bibitem[{{Liu} {et~al.}(2019){Liu}, {Dipierro}, {Ragusa}, {Lodato}, {Herczeg},
  {Long}, {Harsono}, {Boehler}, {Menard}, {Johnstone}, {Pascucci}, {Pinilla},
  {Salyk}, {van der Plas}, {Cabrit}, {Fischer}, {Hendler}, {Manara}, {Nisini},
  {Rigliaco}, {Avenhaus}, {Banzatti}, \& {Gully-Santiago}}]{Liu2019}
{Liu}, Y., {Dipierro}, G., {Ragusa}, E., {et~al.} 2019, \aap, 622, A75,
  \dodoi{10.1051/0004-6361/201834157}

\bibitem[{{Long} {et~al.}(2018){Long}, {Pinilla}, {Herczeg}, {Harsono},
  {Dipierro}, {Pascucci}, {Hendler}, {Tazzari}, {Ragusa}, {Salyk}, {Edwards},
  {Lodato}, {van de Plas}, {Johnstone}, {Liu}, {Boehler}, {Cabrit}, {Manara},
  {Menard}, {Mulders}, {Nisini}, {Fischer}, {Rigliaco}, {Banzatti}, {Avenhaus},
  \& {Gully-Santiago}}]{Long2018}
{Long}, F., {Pinilla}, P., {Herczeg}, G.~J., {et~al.} 2018, \apj, 869, 17,
  \dodoi{10.3847/1538-4357/aae8e1}

\bibitem[{{Lyra} {et~al.}(2010){Lyra}, {Paardekooper}, \& {Mac Low}}]{Lyra2010}
{Lyra}, W., {Paardekooper}, S.-J., \& {Mac Low}, M.-M. 2010, \apjl, 715, L68,
  \dodoi{10.1088/2041-8205/715/2/L68}

\bibitem[{{Mac{\'\i}as} {et~al.}(2018){Mac{\'\i}as}, {Espaillat}, {Ribas},
  {Schwarz}, {Anglada}, {Osorio}, {Carrasco-Gonz{\'a}lez}, {G{\'o}mez}, \&
  {Robinson}}]{Macias2018}
{Mac{\'\i}as}, E., {Espaillat}, C.~C., {Ribas}, {\'A}., {et~al.} 2018, \apj,
  865, 37, \dodoi{10.3847/1538-4357/aad811}

\bibitem[{{Mawet} {et~al.}(2012){Mawet}, {Absil}, {Montagnier}, {Riaud},
  {Surdej}, {Ducourant}, {Augereau}, {R{\"o}ttinger}, {Girard}, {Krist}, \&
  {Stapelfeldt}}]{Mawet2012}
{Mawet}, D., {Absil}, O., {Montagnier}, G., {et~al.} 2012, \aap, 544, A131,
  \dodoi{10.1051/0004-6361/201219662}

\bibitem[{{Meijerink} {et~al.}(2009){Meijerink}, {Pontoppidan}, {Blake},
  {Poelman}, \& {Dullemond}}]{Meijerink2009}
{Meijerink}, R., {Pontoppidan}, K.~M., {Blake}, G.~A., {Poelman}, D.~R., \&
  {Dullemond}, C.~P. 2009, \apj, 704, 1471,
  \dodoi{10.1088/0004-637X/704/2/1471}

\bibitem[{{Menu} {et~al.}(2015){Menu}, {van Boekel}, {Henning}, {Leinert},
  {Waelkens}, \& {Waters}}]{Menu2015}
{Menu}, J., {van Boekel}, R., {Henning}, T., {et~al.} 2015, \aap, 581, A107,
  \dodoi{10.1051/0004-6361/201525654}

\bibitem[{{Mulders}(2018)}]{Mulders2018}
{Mulders}, G.~D. 2018, {Planet Populations as a Function of Stellar
  Properties}, 153, \dodoi{10.1007/978-3-319-55333-7_153}

\bibitem[{{Najita} {et~al.}(2013){Najita}, {Carr}, {Pontoppidan}, {Salyk}, {van
  Dishoeck}, \& {Blake}}]{Najita2013}
{Najita}, J.~R., {Carr}, J.~S., {Pontoppidan}, K.~M., {et~al.} 2013, \apj, 766,
  134, \dodoi{10.1088/0004-637X/766/2/134}

\bibitem[{{Oberg} {et~al.}(2021){Oberg}, {Guzman}, {Walsh}, {Aikawa}, {Bergin},
  {Law}, {Loomis}, {Alarcon}, {Andrews}, {Bae}, {Bergner}, {Boehler}, {Booth},
  {Bosman}, {Calahan}, {Cataldi}, {Cleeves}, {Czekala}, {Furuya}, {Huang},
  {Ilee}, {Kurtovic}, {Le Gal}, {Liu}, {Long}, {Menard}, {Nomura}, {Perez},
  {Qi}, {Schwarz}, {Sierra}, {Teague}, {Tsukagoshi}, {Yamato}, {van 't Hoff},
  {Waggoner}, {Wilner}, \& {Zhang}}]{oberg20}
{Oberg}, K.~I., {Guzman}, V.~V., {Walsh}, C., {et~al.} 2021, arXiv e-prints,
  arXiv:2109.06268.
\newblock \doarXiv{2109.06268}

\bibitem[{{Pi{\'e}tu} {et~al.}(2007){Pi{\'e}tu}, {Dutrey}, \&
  {Guilloteau}}]{Pietu2007}
{Pi{\'e}tu}, V., {Dutrey}, A., \& {Guilloteau}, S. 2007, \aap, 467, 163,
  \dodoi{10.1051/0004-6361:20066537}

\bibitem[{{Pinte} {et~al.}(2018){Pinte}, {M{\'e}nard}, {Duch{\^e}ne}, {Hill},
  {Dent}, {Woitke}, {Maret}, {van der Plas}, {Hales}, {Kamp}, {Thi}, {de
  Gregorio-Monsalvo}, {Rab}, {Quanz}, {Avenhaus}, {Carmona}, \&
  {Casassus}}]{Pinte2018}
{Pinte}, C., {M{\'e}nard}, F., {Duch{\^e}ne}, G., {et~al.} 2018, \aap, 609,
  A47, \dodoi{10.1051/0004-6361/201731377}

\bibitem[{{Pontoppidan} {et~al.}(2008){Pontoppidan}, {Blake}, {van Dishoeck},
  {Smette}, {Ireland}, \& {Brown}}]{Pontoppidan2008spectro}
{Pontoppidan}, K.~M., {Blake}, G.~A., {van Dishoeck}, E.~F., {et~al.} 2008,
  \apj, 684, 1323, \dodoi{10.1086/590400}

\bibitem[{{Pontoppidan} \& {Blevins}(2014)}]{Pontoppidan2014}
{Pontoppidan}, K.~M., \& {Blevins}, S.~M. 2014, Faraday Discussions, 169, 49,
  \dodoi{10.1039/C3FD00141E}

\bibitem[{{Pontoppidan} {et~al.}(2010){Pontoppidan}, {Salyk}, {Blake},
  {Meijerink}, {Carr}, \& {Najita}}]{Pontoppidan2010}
{Pontoppidan}, K.~M., {Salyk}, C., {Blake}, G.~A., {et~al.} 2010, \apj, 720,
  887, \dodoi{10.1088/0004-637X/720/1/887}

\bibitem[{{Rosenfeld} {et~al.}(2012){Rosenfeld}, {Qi}, {Andrews}, {Wilner},
  {Corder}, {Dullemond}, {Lin}, {Hughes}, {D'Alessio}, \& {Ho}}]{Rosenfeld2012}
{Rosenfeld}, K.~A., {Qi}, C., {Andrews}, S.~M., {et~al.} 2012, \apj, 757, 129,
  \dodoi{10.1088/0004-637X/757/2/129}

\bibitem[{{Rosotti} {et~al.}(2021){Rosotti}, {Ilee}, {Facchini}, {Tazzari},
  {Booth}, {Clarke}, \& {Kama}}]{Rosotti2021}
{Rosotti}, G.~P., {Ilee}, J.~D., {Facchini}, S., {et~al.} 2021, \mnras, 501,
  3427, \dodoi{10.1093/mnras/staa3869}

\bibitem[{{Salyk} {et~al.}(2011){Salyk}, {Pontoppidan}, {Blake}, {Najita}, \&
  {Carr}}]{Salyk2011}
{Salyk}, C., {Pontoppidan}, K.~M., {Blake}, G.~A., {Najita}, J.~R., \& {Carr},
  J.~S. 2011, \apj, 731, 130, \dodoi{10.1088/0004-637X/731/2/130}

\bibitem[{{Sch\"{o}ier} {et~al.}(2005){Sch\"{o}ier}, {van der Tak}, {van
  Dishoeck}, \& {Black}}]{Schoier2005}
{Sch\"{o}ier}, F.~L., {van der Tak}, F.~F.~S., {van Dishoeck}, E.~F., \&
  {Black}, J.~H. 2005, \aap, 432, 369, \dodoi{10.1051/0004-6361:20041729}

\bibitem[{{Sierra} {et~al.}(2021){Sierra}, {P{\'e}rez}, {Zhang}, {Law},
  {Guzm{\'a}n}, {Qi}, {Bosman}, {{\"O}berg}, {Andrews}, {Long}, {Teague},
  {Booth}, {Walsh}, {Wilner}, {M{\'e}nard}, {Cataldi}, {Czekala}, {Bae},
  {Huang}, {Bergner}, {Ilee}, {Benisty}, {Le Gal}, {Loomis}, {Tsukagoshi},
  {Liu}, {Yamato}, \& {Aikawa}}]{sierra20}
{Sierra}, A., {P{\'e}rez}, L.~M., {Zhang}, K., {et~al.} 2021, arXiv e-prints,
  arXiv:2109.06433.
\newblock \doarXiv{2109.06433}

\bibitem[{{Simon} {et~al.}(2019){Simon}, {Guilloteau}, {Beck}, {Chapillon}, {Di
  Folco}, {Dutrey}, {Feiden}, {Grosso}, {Pi{\'e}tu}, {Prato}, \&
  {Schaefer}}]{Simon2019}
{Simon}, M., {Guilloteau}, S., {Beck}, T.~L., {et~al.} 2019, \apj, 884, 42,
  \dodoi{10.3847/1538-4357/ab3e3b}

\bibitem[{{Teague} {et~al.}(2019){Teague}, {Bae}, \& {Bergin}}]{Teague2019}
{Teague}, R., {Bae}, J., \& {Bergin}, E.~A. 2019, \nat, 574, 378,
  \dodoi{10.1038/s41586-019-1642-0}

\bibitem[{{van der Plas} {et~al.}(2015){van der Plas}, {van den Ancker},
  {Waters}, \& {Dominik}}]{vanderPlas2015}
{van der Plas}, G., {van den Ancker}, M.~E., {Waters}, L.~B.~F.~M., \&
  {Dominik}, C. 2015, \aap, 574, A75, \dodoi{10.1051/0004-6361/201425052}

\bibitem[{{Weaver} {et~al.}(2018){Weaver}, {Isella}, \& {Boehler}}]{Weaver2018}
{Weaver}, E., {Isella}, A., \& {Boehler}, Y. 2018, \apj, 853, 113,
  \dodoi{10.3847/1538-4357/aaa481}

\bibitem[{{Wilson} \& {Rood}(1994)}]{Wilson1994}
{Wilson}, T.~L., \& {Rood}, R. 1994, \araa, 32, 191,
  \dodoi{10.1146/annurev.aa.32.090194.001203}

\bibitem[{{Zhang} {et~al.}(2017){Zhang}, {Bergin}, {Blake}, {Cleeves}, \&
  {Schwarz}}]{zhang2017}
{Zhang}, K., {Bergin}, E.~A., {Blake}, G.~A., {Cleeves}, L.~I., \& {Schwarz},
  K.~R. 2017, Nature Astronomy, 1, 0130, \dodoi{10.1038/s41550-017-0130}

\bibitem[{{Zhang} {et~al.}(2020{\natexlab{a}}){Zhang}, {Bosman}, \&
  {Bergin}}]{Zhang2020a}
{Zhang}, K., {Bosman}, A.~D., \& {Bergin}, E.~A. 2020{\natexlab{a}}, \apjl,
  891, L16, \dodoi{10.3847/2041-8213/ab77ca}

\bibitem[{{Zhang} {et~al.}(2020{\natexlab{b}}){Zhang}, {Schwarz}, \&
  {Bergin}}]{Zhang2020b}
{Zhang}, K., {Schwarz}, K.~R., \& {Bergin}, E.~A. 2020{\natexlab{b}}, \apjl,
  891, L17, \dodoi{10.3847/2041-8213/ab7823}

\bibitem[{{Zhang} {et~al.}(2021){Zhang}, {Booth}, {Law}, {Bosman}, {Schwarz},
  {Bergin}, {{\"O}berg}, {Andrews}, {Guzm{\'a}n}, {Walsh}, {Qi}, {van 't Hoff},
  {Long}, {Wilner}, {Huang}, {Czekala}, {Ilee}, {Cataldi}, {Bergner}, {Aikawa},
  {Teague}, {Bae}, {Loomis}, {Calahan}, {Alarc{\'o}n}, {M{\'e}nard}, {Le Gal},
  {Sierra}, {Yamato}, {Nomura}, {Tsukagoshi}, {P{\'e}rez}, {Trapman}, {Liu}, \&
  {Furuya}}]{zhang20}
{Zhang}, K., {Booth}, A.~S., {Law}, C.~J., {et~al.} 2021, arXiv e-prints,
  arXiv:2109.06233.
\newblock \doarXiv{2109.06233}

\bibitem[{{Zhu} {et~al.}(2019){Zhu}, {Zhang}, {Jiang}, {Kataoka}, {Birnstiel},
  {Dullemond}, {Andrews}, {Huang}, {P{\'e}rez}, {Carpenter}, {Bai}, {Wilner},
  \& {Ricci}}]{Zhu2019}
{Zhu}, Z., {Zhang}, S., {Jiang}, Y.-F., {et~al.} 2019, \apjl, 877, L18,
  \dodoi{10.3847/2041-8213/ab1f8c}

\end{thebibliography}

\newpage
\appendix
\section{Line profiles}

\subsection{Asymmetries in the CO $J$=2--1 line profiles}
\label{ssc:linesym}
\begin{figure*}
    \centering
    \includegraphics[width=\hsize]{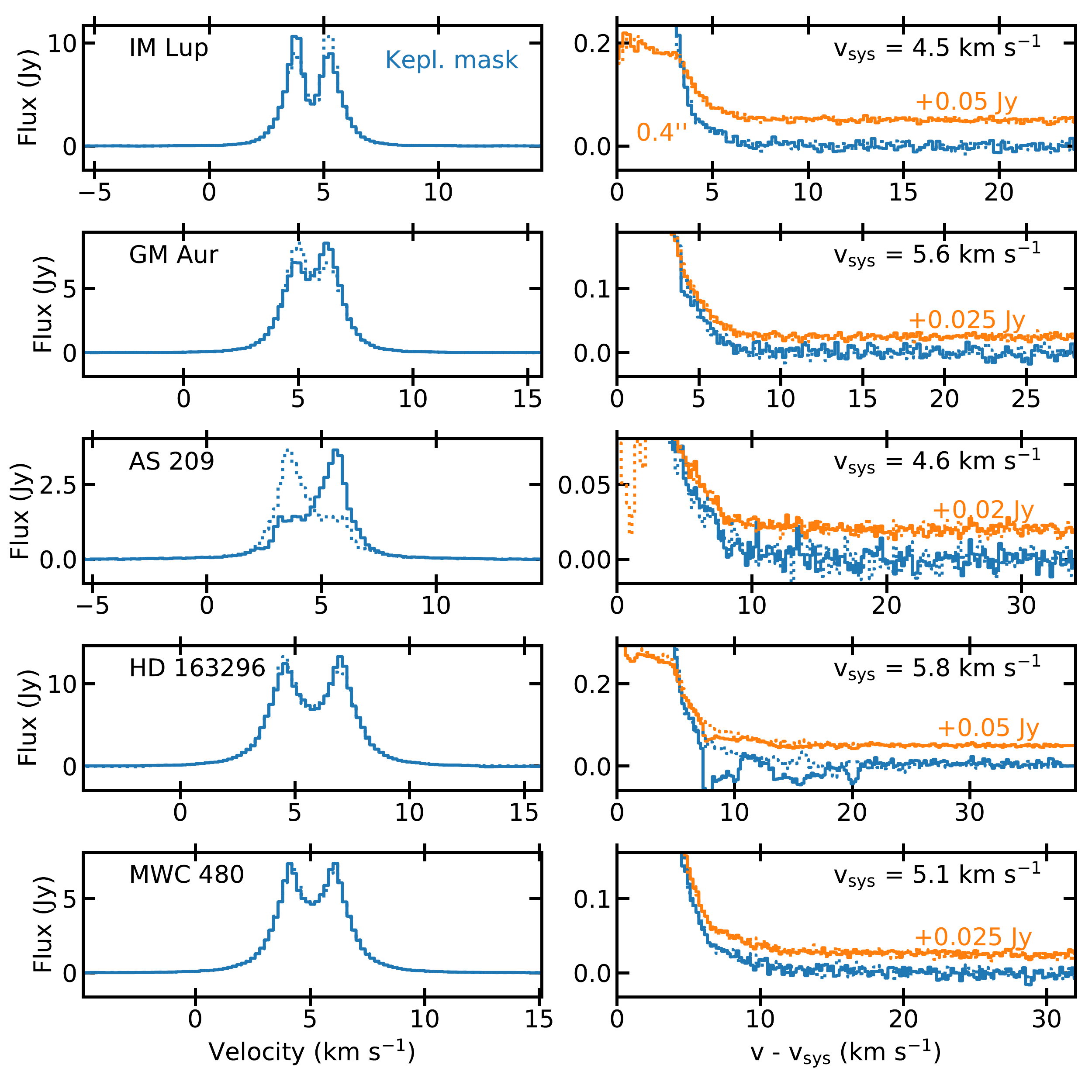}
    \caption{\ce{^{12}CO} $J$=2--1 spectra for the five MAPS sources. Spectra in blue are extracted by integrating over the cleaning mask in each channel. Each spectrum has been mirrored around the source velocity; this is plotted in dotted lines. The right column shows a zoom of these spectra, plotted against the offset from source velocity. The red side is shown with solid lines, while the dashed lines show the blue side. In orange a spectrum extracted with a smaller circular aperture is shown, these are offset in flux for clarity. In the left column, some clear asymmetries in the line profiles at small velocity offsets can be seen, either due to large scale disk structures (GM Aur, IM Lup and HD 163296) or foreground emission (AS 209). The spectra for GM Aur, IM Lup and MWC 480 at high velocities, hence small radii, are very symmetric. HD 163296 shows significant asymmetry at large velocity offsets and AS 209 shows an anomalously high flux on the red side at 5.9 km s$^{-1}$ offset (10.5 km s$^{-1}$ in local standard of rest). These feature are shown in more detail in Fig.~\ref{fig:zoomed_spectra}.   }
    \label{fig:asym_12CO}
\end{figure*}

\begin{figure}
    \centering
    \includegraphics[width=\hsize]{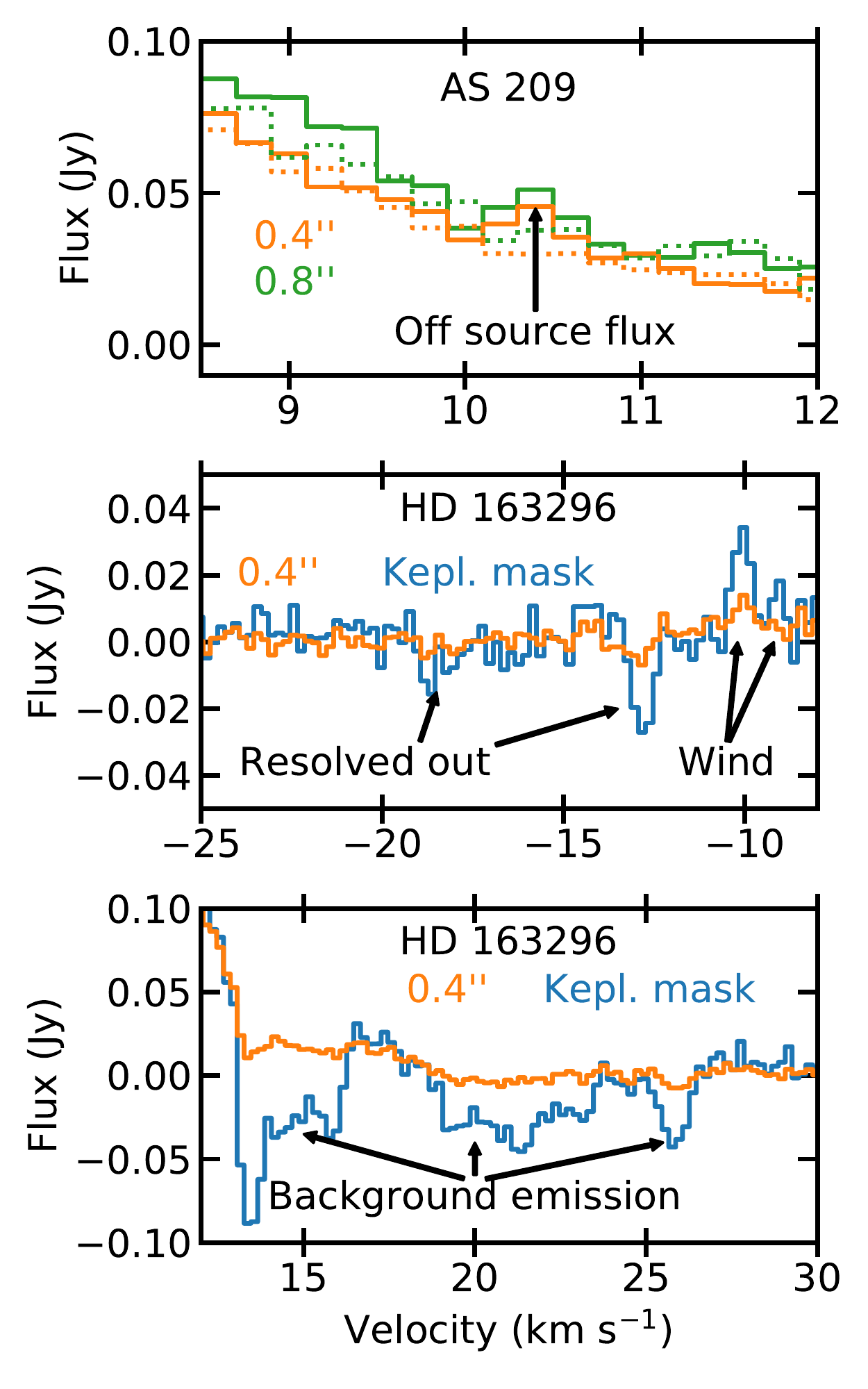}
    \caption{Zoom-in of the \ce{^{12}CO} $J$=2--1 spectra for the asymmetric features in AS 209 and HD 163296. Spectra in blue (HD 163296 only) are extracted by integrating over the cleaning mask in each channel, the orange spectra from a circular aperture of 0\farcs4 and the green spectra (AS 209 only) from a circular aperture of 0\farcs8. The AS 209 spectrum also shows the mirror spectrum in green. All features have been labeled. The flux spike in AS 209 (top), most clearly seen in the 0\farcs8 spectrum, is assumed to be due to a noise spike. The excess emission on the blue size of HD 163296 (middle) is due to wind emission falling within the Keplerian mask, whereas the negative fluxes are due to the fringing caused by the strong and extended wind emission. On the red side of HD 163296 (bottom), strong negative fluxes are measured in the Keplerian mask as the disk is backlit by galactic CO emission at these velocities, creating negatives in the continuum subtracted images where there is dust emission.}
    \label{fig:zoomed_spectra}
\end{figure}

To extract the radial profiles, we assume that all emission originates from a flat Keplerian disk. While it is impossible to ascertain if emission is actually coming from a Keplerian disk without resolving the emission, it is possible to look for non-Keplerian gas by looking at the asymmetries in the line emission. Where-ever this is symmetry broken, line emission on either the blue- or red-shifted side of the line is certainly not coming from a Keplerian disk. These asymmetries should be most obvious in the \ce{^{12}CO} spectra.

Fig.~\ref{fig:asym_12CO} shows the \ce{^{12}CO} $J$=2--1 spectra and their mirror image, in velocity space, for the 5 MAPS disks. The systematic velocities as tabulated in Table~\ref{tab:mass_incl} are used to mirror the line profiles. At velocities close to the systemic velocity in Figure~\ref{fig:asym_12CO}, differences between the low velocity red and blue shifted emission can be seen for AS 209, IM Lup, GM Aur, and HD 163296. These line asymmetries can all be ascribed to factors discussed elsewhere \citep[foreground absorption, non-disk emission, disk flaring,][]{law20_rad, huang20}. The line wings for IM Lup, GM Aur and MWC 480 show very symmetric emission, suggesting that a Keplerian disk is a good approximation. AS 209 and HD 163296 show features that warrant further investigation. Regions of interest are plotted in Fig.~\ref{fig:zoomed_spectra}.

AS 209 shows a feature around 10.5 km s$^{-1}$. There is a 2 channel, 6-7 sigma flux spike that is offset from the source, and is thus stronger in the 0\farcs8 aperture than the 0\farcs4 aperture (see Fig.~\ref{fig:zoomed_spectra}). The image cubes show a compact emission component $\sim$ 0\farcs3 offset from the disk center. This component does not have a counterpart in \ce{^{13}CO} $J$=2--1, nor in the DSHARP $J$=2--1 \ce{^{12}CO} datacubes \citep{Andrews2018}. The feature is considered anomalous and the channels around this region are masked in the rest of the analysis.

The spectrum of HD 163296 is very messy at high velocity offsets. On the blue side of source velocity there is a known CO outflow \citep[][]{Klaassen2013, booth2020}. Around -10 km s$^{-1}$ some of the emission from the outflow is contaminating our on source spectra, especially in the Keplerian CLEANing mask\footnote{The CLEANing mask at these velocities is large enough to capture the size of the dust disk and is thus about 3\arcsec{} across.}. Between -12 and -20 km s$^{-1}$ there are some negative flux features in the spectrum. The velocities of these correspond to velocities at which the outflow is bright and extended. As the full extend of the outflow is not properly captured in the interferometric data, this induces a fringe pattern over the entire reconstructed image, leading to negative fluxes at some velocities. As both flux contribution due to the outflow as well as the modification to the flux due the fringing is hard to quantify, all channel with velocities between -25 and -9 km s$^{-1}$ are excluded from further analysis

On the red side of the spectrum of HD 163296, strong negative fluxes can be seen in the spectrum extracted within the Keplerian CLEANing mask between 13 and 26 km s$^{-1}$. The image cubes show evidence of strong CO emission behind the disk at these velocities. This back-lighting of the disk leads to the dust disk showing up as a negative in these continuum subtracted images and the large scale nature of the background emission leads to strong fringing, further modifying the flux. Again we have decided to remove these velocities from further analysis. As both the red and blue side of the spectrum are unusable at high velocities, our radial profile extraction is limited to radii larger than 5 au in HD 163296. Future observations, with better short baseline spacing should be able to resolve some of these problems allowing to extract an accurate inner disk CO flux from these high velocity channels.

\subsection{\ce{^{13}CO} $J$=2--1 and $J$=1--0 comparison}

\begin{figure*}
    \centering
    \includegraphics[width=\hsize]{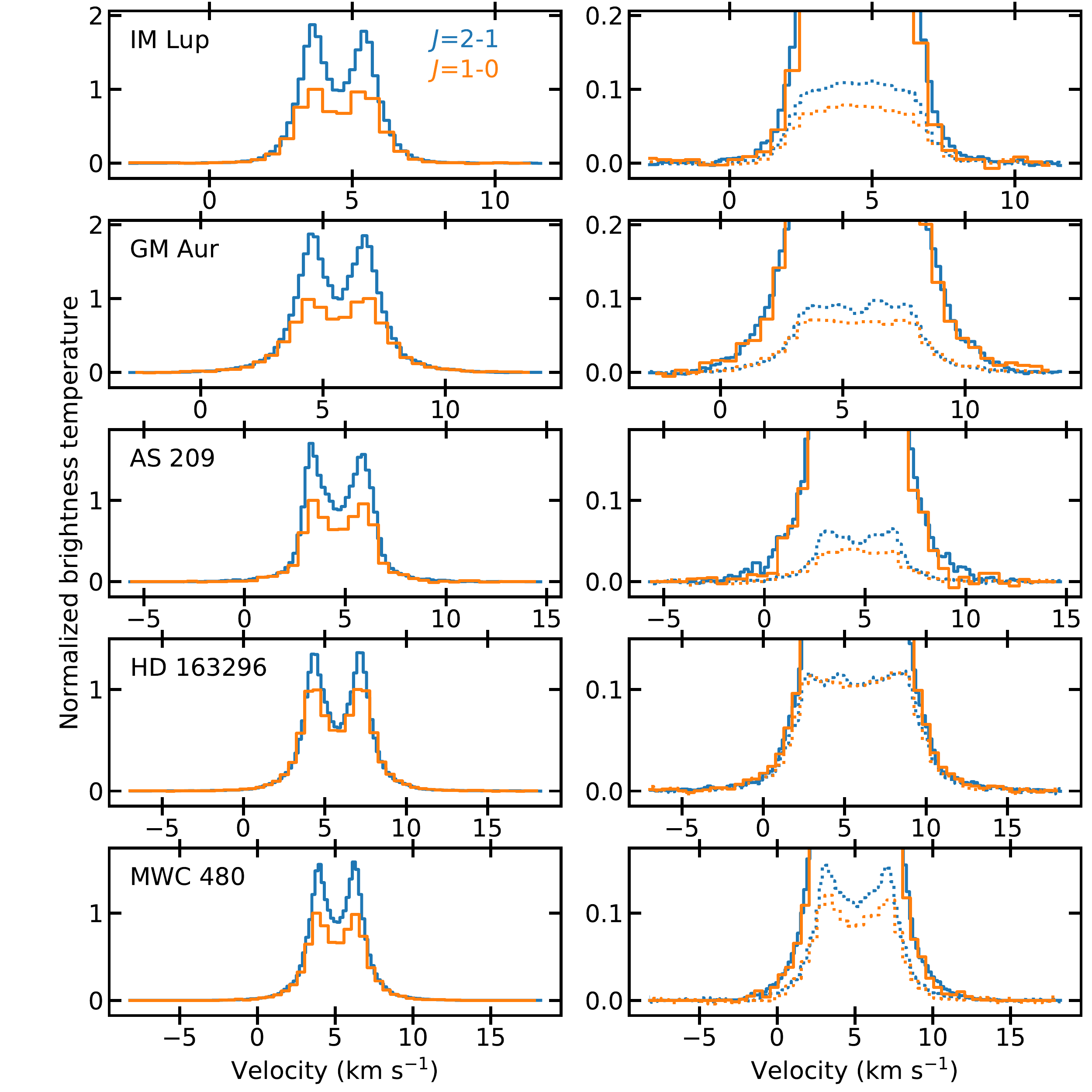}
    \caption{Comparison of the \ce{^{13}CO} $J$=2--1 (blue) and \ce{^{13}CO} $J$=1--0 (orange) line profiles. The profiles are normalized to the $J$=1--0 line. The left figure shows the full line profile, while the right shows a more constrained brightness temperature range, with spectra extracted with a 1\farcs0 aperture as dotted lines. The profiles are scaled to a brightness temperature scale assuming Rayleigh–Jeans law, which underestimates the $J$= 2--1 line strength (see Sec.~\ref{ssc:band_rat}). For velocities close to the systemic value, the $J$= 2--1 line is consistently brighter than the $J$=1--0 line, indicating that the \ce{^{13}CO} flux is either optically thin, or that the $J$=2--1 flux comes from a warmer layer than the $J$= 1--0. In the line wings however, it seems that both lines are equally bright. This indicates that both the $J$=2--1 and the $J$=1--0 lines are optically thick.}
    \label{fig:B3vsB6}
\end{figure*}

\begin{figure}
    \centering
    \includegraphics[width=\hsize]{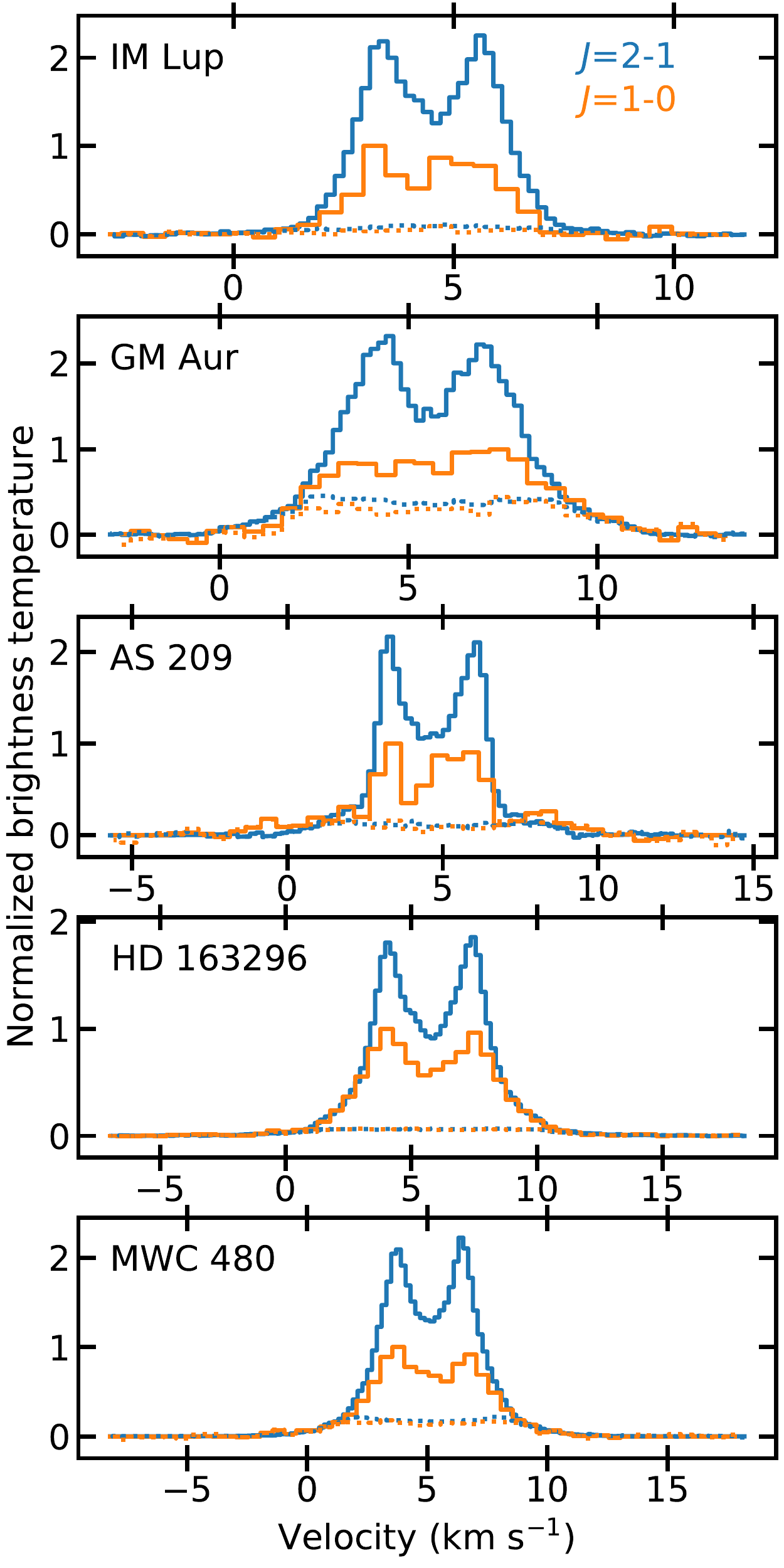}
    \caption{Comparison of the \ce{C^{18}O} $J$=2--1 (blue) and \ce{C^{18}O} $J$=1--0 (orange) line profiles, solid lines shows the profiles extracted with the Keplerian cleaning mask and the spectrum extracted with a 0\farcs4 aperture is shown in dotted lines. The profiles are scaled to a brightness temperature scale assuming Rayleigh–Jeans law, which underestimates the $J$= 2--1 line strength (see Sec.~\ref{ssc:band_rat}). Similar to Fig.~\ref{fig:B3vsB6}, no evidence of stronger emission in Band 3 is seen compared to Band 6 in the line wings.}
    \label{fig:B3vsB6_18}
\end{figure}
\label{ssc:band_rat}
In the absence of a dust cavity, dust in the inner regions of protoplanetary disks is so abundant that it should be optically thick at millimeter wavelengths. It is generally assumed that due to the lower opacity at longer wavelength observations at longer wavelength can penetrate significantly deeper into the disk, leading to brighter (in units of Kelvin) lines.
The difference in dust extinction opacity between 0.3 and 3 mm (approximately the wavelength range currently covered by ALMA) is assumed to be less than an order of magnitude in a standard large grain (sizes up to 1 millimeter) dust models \citep[e.g.][]{Birnstiel2018}. This suggests that the differences in probed column and line fluxes might be small. The MAPS data covers both the $J$=2--1 (220 GHz) and the $J$=1--0 (110 GHz)\ce{^{13}CO} lines at high resolution ($<$0.3\arcsec) and sensitivity. Higher probing depth, due to the lower dust opacity, of the lower frequency line should show in source spectra as excess line wing flux, when compared on equal scales. The Band 3 (3mm, $J$=1--0) and Band 6 (1.3mm, $J$=2.1) line profiles of \ce{^{13}CO} and \ce{C^{18}O} are compared in Figs.~\ref{fig:B3vsB6}~and~\ref{fig:B3vsB6_18} respectively. The $J$=2--1 lines are scaled down by a factor 4 to compensate for the Rayleigh-Jeans relation, $I\propto \nu^2 T$. Compared to the full Planck law this underestimates the $J$=2--1 flux by 10\% at 30 K and with values agreeing better at higher temperatures.  

Figures~\ref{fig:B3vsB6}~and~\ref{fig:B3vsB6_18} show that, near the systemic velocity, the $J$=2--1 line is consistently brighter than the $J$=1--0 line. This can be understood from the Einstein A coefficients and molecular level structure, which predicts an optical depth ratio (and thus brightness temperature ratio if both lines are optically thin) between the $J$=2--1 and $J$=1--0 lines of 2.7-3 for gas of 20-30 K \citep{Schoier2005, Endres2016}. These ratios are not reached implying a strong contribution from optically thick $J$=2--1 line flux even low velocities relative to source. In the line wings, however, the $J$=2--1 and $J$=1--0 spectra have indistinguishable emission temperatures, as expected for optically thick lines with the same $T_\mathrm{ex}$. In the inner $\sim$ 40 au the $J=$1--0 and $J$=2--1 line thus originate for a very similar layer. This indicates that the lower dust opacity at Band 3 is not allowing the observations to probe deeper into the disk compared to higher frequency bands. 

\section{Surface brightness fitting process}
\label{app:fitting}

\begin{figure}
    \centering
    \includegraphics[width = \hsize]{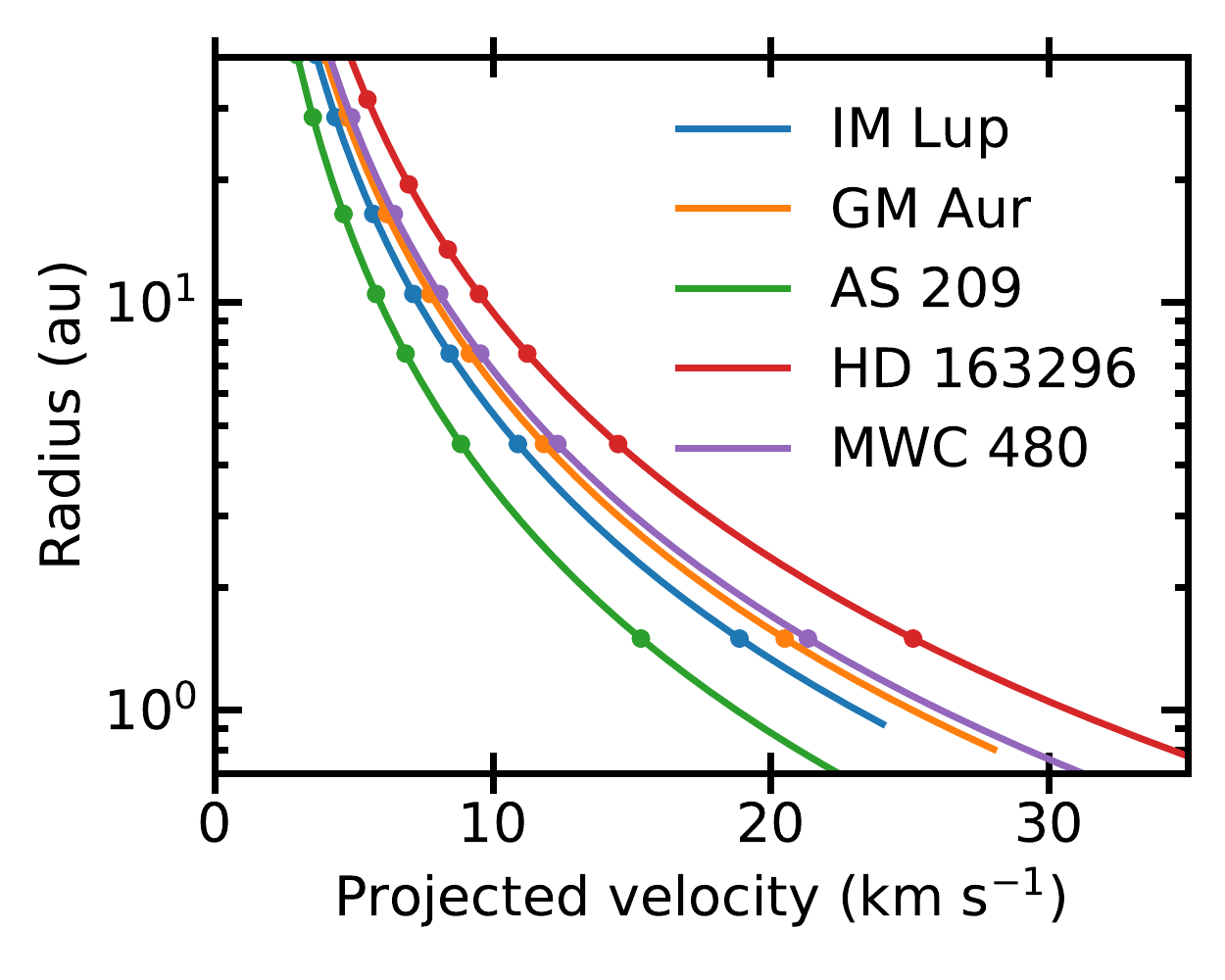}
    \caption{Relation between projected velocity offset and Keplerian radius (Eq.~\ref{eq:v_r_kepl}) as well as the radial grid points used in the fitting procedure, assuming a minimum of 3 velocity bins per radial bin. }
    \label{fig:r_vs_v}
\end{figure}

\begin{figure}
    \centering
    \includegraphics[width = \hsize]{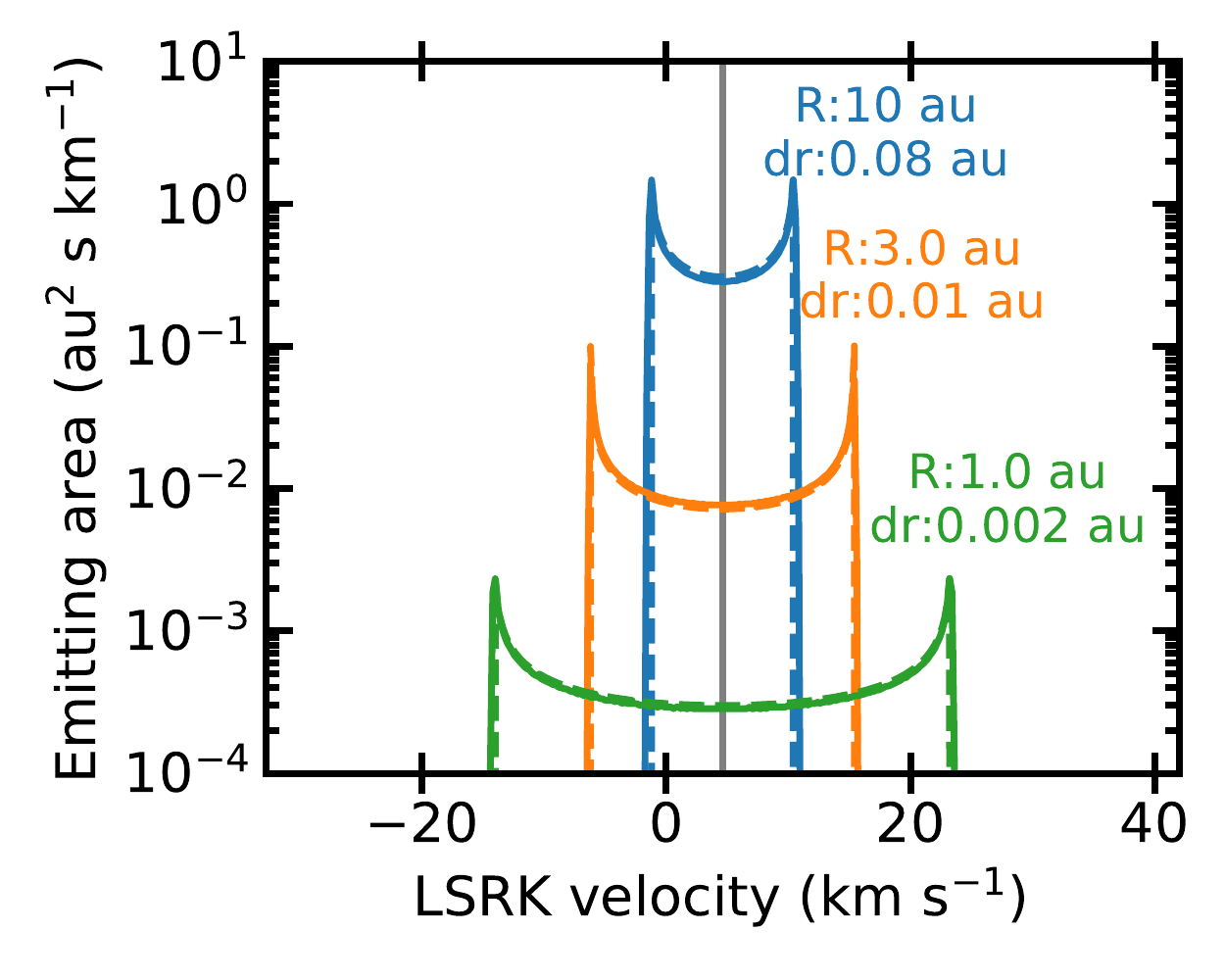}
    \caption{Shape and strength of three components used in the fitting procedure of the AS 209 disk (solid lines) compared to the analytic shape (Eq.~\ref{eq:kepl_shape}, dashed lines). The components have been convolved with a Gaussian of 0.2~km~s$^{-1}$, which accounts for the slight differences from the analytic shape. For each, component both the radius, as well as the width of the annulus it represents are given.}
    \label{fig:comp}
\end{figure}

\begin{figure*}
    \centering
    \includegraphics[width = \hsize]{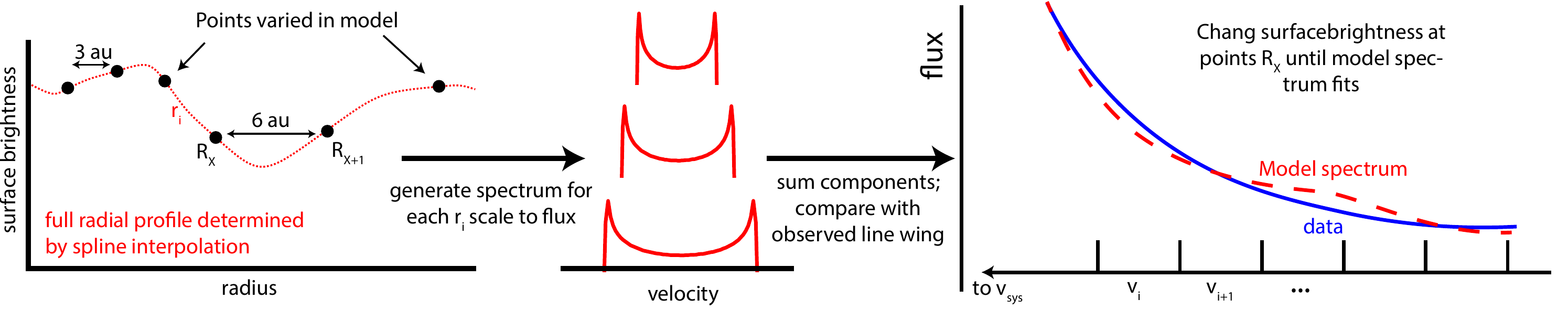}
    \caption{Schematic representation of the fitting process. The left panel shows the input points, R$_X$, in black. These points have semi-regular spacing of 3, 6 and 12 au (see Fig.~\ref{fig:r_vs_v}) for the radial locations of these points. From the surface brightness at points R$_X$, the surface brightness at a secondary grid, r$_i$ is calculated by spline interpolation. The points r$_i$ are spaced in radius such that their $v_\mathrm{max}(\mathrm{r}_i)$ are regularly spaced in velocity with a 0.025 km s$^{-1}$ spacing. For each of these radius r$_i$ the spectrum of a annulus of this radius is calculated and scaled to the required intensity (middle). These spectral components are summed to calculated the full line profile. This is then compared to the observed line profile (right). The intensity at the points R$_X$ is then varied until there is a match between observed and modeled spectrum.}
    \label{fig:schem_fit}
\end{figure*}

The fitting process to extract radial surface brightness profiles from the line profile wings is given schematically in Fig.~\ref{fig:schem_fit}. The fitting procedure starts with creating the forward model components which have a Keplerian line shape, corresponding to an annulus in the disk at a given radius. The shape is given by:
\begin{equation}
    \label{eq:kepl_shape}
    f(v) = \frac{1}{\sqrt{v^2_\mathrm{max} - v^2}},
\end{equation}
with $v_\mathrm{max}$ as in Eq.~\ref{eq:v_r_kepl}. These components are convolved with a 0.2 km s$^{-1}$ FWHM gaussian to account for instrumental and physical broadening of the line and then normalized to their surface area, which is influenced by the spacing between flux components, examples of components for the AS 209 disk are shown in Fig.~\ref{fig:comp}.

To correctly fit the line profiles wings, the underlying emission model needs to have enough resolution to fully represent the disk emission. Given that the spectra are not Nyquist sampled and that the sampling is smaller than the expected line width\footnote{Only for $T>$100 K is the thermal linewidth larger than the 0.2 km s$^{-1}$ velocity resolution}, the velocity spacing of the components has to be smaller than the bin size. Here we use a spacing that is 8 times smaller, leading to model components (r$_\mathrm{i}$ in Fig.~\ref{fig:schem_fit}) being spaced by 0.025 km$^{-1}$, and irregularly spaced in radius. This leads to 500-2000 components for a single line. This is far too many to fit individually. Therefore a lower resolution grid is made (R$_\mathrm{X}$ in Fig.~\ref{fig:schem_fit}), the intensity at these radii are varied as the free parameters to the fitting problem, the intensity of the model components that actually make up the final spectrum is derived from these by spline interpolation.

To leverage the sensitivity of this technique at small radii, this grid is irregular in radius space as well. The fitting points are initially spaced by 3 au, when the projected Keplerian velocity difference between two points becomes less than 0.6 km s$^{-1}$, the grid spacing is increased to 6 au, and when this is still not enough to 12 au. As such our radial surface brightness profiles have a lower resolution at larger radii. The radii at which these points fall is shown in Fig.~\ref{fig:r_vs_v}.

\section{Comparison with image radial profiles}
\label{app:image_comp}
\begin{figure}
    \centering
    \includegraphics[width=\hsize]{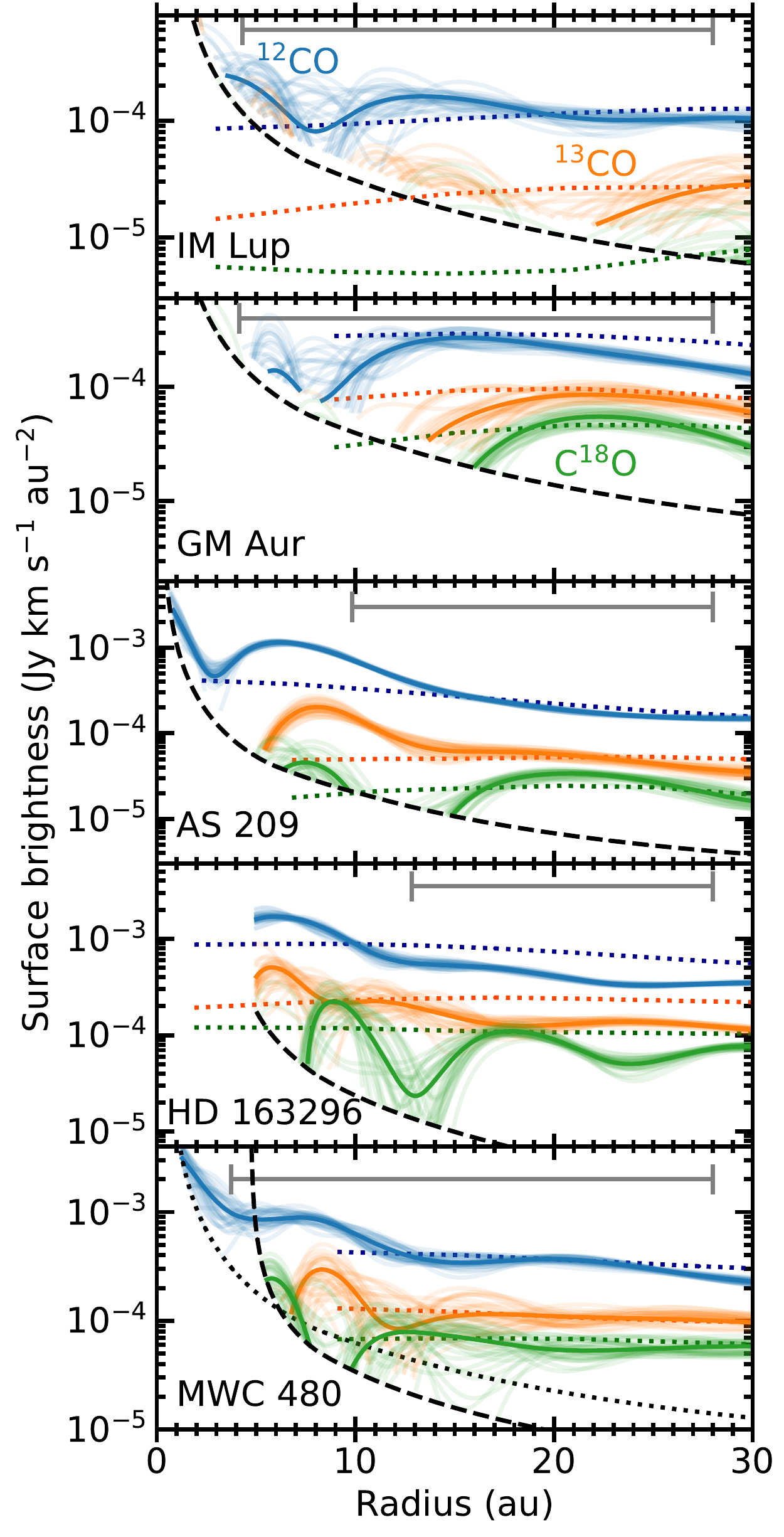}
    \caption{Same as Fig.~\ref{fig:Radial_profiles}, with the image cube extracted radial profiles \citep[dotted][]{law20_rad}. Grey bars, show the nominal MAPS resolution of 0\farcs15. Image radial profile extraction is done on the fiducial 0\farcs15 resolution MAPS images with a 15$^{\circ}$ wedge for all lines except the \ce{^{12}CO} line in AS 209, where a one sided 55$^{\circ}$ wedge is used \citep[see][no vertical emission height has been assumed for any of these radial profiles]{law20_rad}  }
    \label{fig:Radial_profiles_w_image}
\end{figure}

Fig.~\ref{fig:Radial_profiles_w_image} shows the line profile extracted surface brightness profiles with the image derived profiles. The profiles agree on the absolute level of flux but the line profile extracted profiles show more and sharper features.

\section{Toy model setup}
\label{app:Toymodels}
To test the effect of large dust in a realistic setting a couple of models using Dust and LInes \citep[DALI, ][]{Bruderer2012, Bruderer2013} have been used. A simple T-Tauri disk motivated by the results from \citet{zhang20} for IM Lup is used for this purpose. Model parameters can be found in Table~\ref{tab:model_param}. Most critically, the gas surface density is chosen such that the CO column in the inner disk is approximately $10^{19}$ cm$^{-2}$ around 40 au, and has a CO abundance of $10^{-5}$, 1 order of magnitude reduced from ISM levels. The disk gas-to-dust ratio is taken to be 100, with 99\% of this dust as large dust settles to the midplane, creating a disk that is only optically thick around 200 GHz within 5 au.

\begin{figure*}[!h]
    \centering
    \includegraphics[width = \hsize]{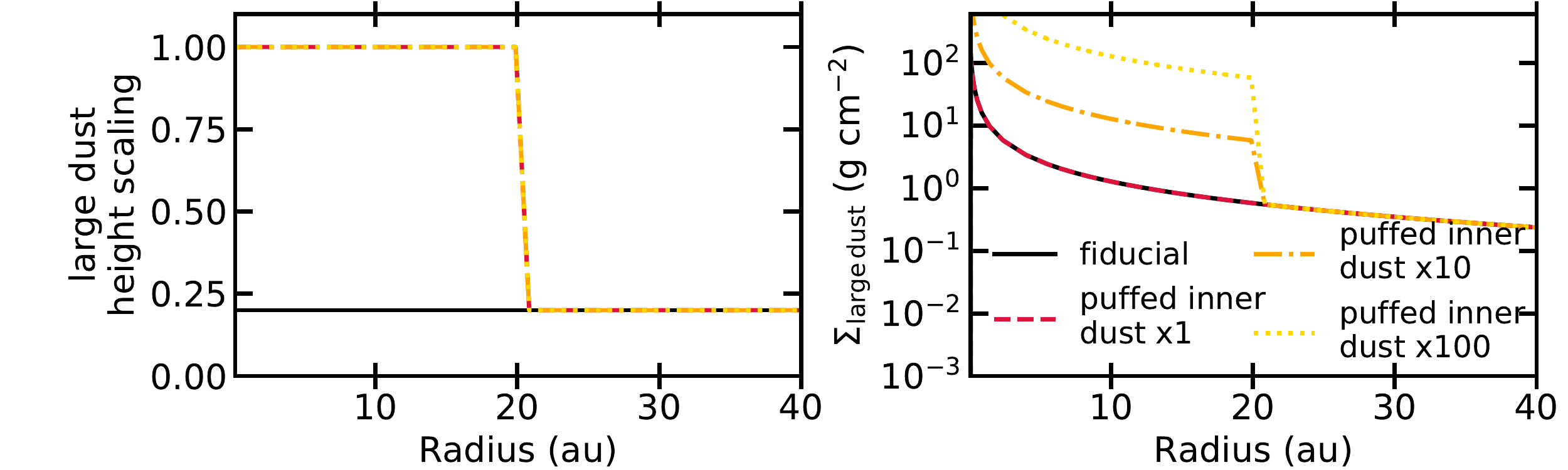}
    \caption{Large dust scale height with respect to the gas scale height (left) and large dust surface density (right) in the DALI models.}
    \label{fig:surfdens}
\end{figure*}

The models vary in the treatment of the large dust within 20 au. For the fiducial model the large dust within 20 au is settled at 20\% of the gas scale-height and the vertically integrated gas-to-dust ratio is 100. The three other models have the dust vertically well mixed with the gas within 20 au. This causes the gas and dust emitting regions to overlap. Furthermore, the vertically integrated gas-to-dust ratio is varied between 1 and 100, by increasing the amount of dust by up to a factor 10. Large dust surface densities and scale-height distribution are shown in Fig.~\ref{fig:surfdens}.

\begin{table}[!h]
\caption{Toy model parameters}
    \centering
    \begin{tabular}{l r}
    \hline
    \hline
    parameter & value \\
    \hline
    $M_\star$  & 1 $M_\odot$ \\
    Stellar spectrum   & IM Lup$^{a}$  \\
    $L_X$ & $4\times 10^{30}$ erg s$^{-1}$ \\
    $\Sigma_\mathrm{c, gas}$ & 28.4 g\,cm$^{-2}$ \\
    $R_c$ & 100 au\\
    $h_c$ & 0.1 \\
    $\gamma$ & 1 \\
    $\psi$ & 0.17 \\
    C/H &$ 1.35\times 10^{-5}$ \\
    O/H &$ 2.88\times 10^{-5}$ \\
    N/H &$ 2.1\times 10^{-6}$ \\
    \hline
    \end{tabular}
    \tablecomments{$^a$ \citet{zhang20}}
    \label{tab:model_param}
\end{table}

\end{document}